# Suite of Meshless Algorithms for Accurate Computation of Soft Tissue Deformation for Surgical Simulation

Grand Joldes[1], George Bourantas[1], Benjamin Zwick[1], Habib Chowdhury[1],
Adam Wittek[1], Sudip Agrawal[1], Konstantinos Mountris[2], Damon Hyde[3],
Simon K. Warfield[3], and Karol Miller[1,4]

[1] Intelligent Systems for Medicine Laboratory, The University of Western Australia, Crawley-Perth 6009, Western Australia, Australia

[2] Aragón Institute for Engineering Research, University of Zaragoza, IIS Aragón, Spain

[3] Computational Radiology Laboratory, Boston Children's Hospital and Harvard Medical School, Boston, Massachusetts, 02115, US.

[4] Institute of Mechanics and Advanced Materials, Cardiff School of Engineering, Cardiff University, Wales, UK

**Abstract:** The ability to predict patient-specific soft tissue deformations is key for computer-integrated surgery systems and the core enabling technology for a new era of personalized medicine. Element-Free Galerkin (EFG) methods are better suited for solving soft tissue deformation problems than the finite element method (FEM) due to their capability of handling large deformation while also eliminating the necessity of creating a complex predefined mesh. Nevertheless, meshless methods based on EFG formulation, exhibit three major limitations: i) meshless shape functions using higher order basis cannot always be computed for arbitrarily distributed nodes (irregular node placement is crucial for facilitating automated discretization of complex geometries); ii) imposition of the Essential Boundary Conditions (EBC) is not straightforward; and, iii) numerical (Gauss) integration in space is not exact as meshless shape functions are not polynomial. This paper presents a suite of Meshless Total Lagrangian Explicit Dynamics (MTLED) algorithms incorporating a Modified Moving Least Squares (MMLS) method for interpolating scattered data both for visualization and for numerical computations of soft tissue deformation, a novel way of imposing EBC for explicit time integration, and an adaptive numerical integration procedure within the Meshless Total Lagrangian Explicit Dynamics algorithm. The appropriateness and effectiveness of the proposed methods is demonstrated using comparisons with the established non-linear procedures from commercial finite element software ABAQUS and experiments with very large deformations. To demonstrate the translational benefits of MTLED we also present a realistic brain-shift computation.

## 1. Introduction

In applications such as surgical simulation or image registration, the key objectives of computational biomechanics are to enable a surgeon to simulate surgery within the operating theatre, using cost-effective and readily-available computing hardware and to visualize the results immediately. A surgeon – the ultimate user of computational biomechanics software, should be able to evaluate the implications of each stage of a surgical procedure and explore potential alternative solutions without requiring any in-depth knowledge of numerical computation. For this purpose, the creation of a straightforward to generate and easily-manipulated patient-specific computational grid, as well as a robust, accurate method for solving the fundamental equations describing the biomechanical behavior (i.e. non-linear partial differential equations of solid mechanics) of the body organs and tissues are the essential requirements.

For the last four decades, Finite Element Analysis (FEA) has been the method of choice in computational biomechanics. Nevertheless, the conventional approach to compute soft tissue deformation depended on linear finite element algorithms which assumed infinitesimal deformations (Cotin et al., 1999; Warfield et al., 2002). However, modeling of soft tissue organs for surgical simulation and image-guided surgery is a non-linear problem of continuum mechanics which involves large deformations and large strains with geometric and material non-linearities (Miller, 2000; Miller, 2011) clearly incompatible with the assumption of infinitesimality of deformations.

Co-rotational finite elements (Crisfield and Moita, 1996) were proposed to allow close-to-real time computation of deformations, however this formulation assumes small strains and linearity of the material response, assumptions clearly not satisfied in many clinically relevant scenarios.

Another difficulty with using the Finite Element Method for patient-specific applications arises from the common practice of using 4-noded tetrahedral (i.e. linear) finite elements. These elements exhibit volumetric locking and should not be used for almost incompressible materials such as soft tissues (Bathe, 1996; Bonet et al., 2001; Hughes, 1987; Joldes et al., 2008b). Parabolic (10-noded) tetrahedron is appropriate but computationally inefficient (Yang, 2018). 8-noded hexahedra are preferable, but efficient generation of hexahedral meshes for complicated geometries, despite enormous research effort (Carey, 1997), still awaits a satisfactory solution (Wittek et al., 2016).

To allow real-time computation of finite deformations of non-linear soft tissues, (Miller et al., 2007) developed the Total Lagrangian Explicit Dynamics (TLED) finite element algorithm that has become an integral part of our Finite Element suite of algorithms for surgical simulation (Joldes et al., 2009a), and was implemented on Graphics Processing Unit for real time applications (Joldes et al., 2010a). The adoption of TL formulation allows pre-computation of all derivatives with respect to spatial co-ordinates and the explicit time integration based on the central difference method eliminates the necessity for iteration during each time-step (Bathe, 1996). Several applications have been demonstrated in surgical simulation, image registration and injury biomechanics based on this framework (Garlapati et al., 2014; Hu et al., 2012; Li et al., 2015; Madhukar and Ostoja-Starzewski, 2019; Mostayed et al., 2013b; Strbac et al., 2017; Wittek et al., 2010).

Despite its accuracy and computational efficiency, TLED (as any other Finite Element scheme) is very difficult to implement in clinical workflows as it requires a high quality finite element mesh. Creation of such a patient-specific mesh from medical images involves image segmentation, creation of water-tight surfaces from the segmentation and discretization of the complex geometries of body organs defined by these surfaces into interconnected meshes of high-quality elements. These pre-processing steps are very labor intensive and difficult to automate (Wittek et al., 2016). Moreover, the finite element solution accuracy deteriorates (or even fails) when elements undergo distortion under large deformations induced by interactions between the body organs/tissues and surgical tools.

To alleviate these limitations, Meshless Methods (MMs) (Gu, 2005; Li and Liu, 2004; Liu and Gu, 2005) have been suggested as a possible alternative to Finite Element Method. Meshless Local Petrov-Galerkin (MLPG) methods have been extensively researched (Atluri, 2002; Atluri and Zhu, 1998)but as yet compelling examples of the successful application of the method to realistic 3D non-linear problems are not available. Approaches based on Smooth Particle Hydrodynamics (Monaghan, 1992) were recently proposed (Ahmadzadeh et al., 2018) but they are not yet rigorously verified and remain very inefficient computationally when applied to solid mechanics problems of complicated geometries. Recently, we proposed a very different meshless approach based on strong form formulation of solid mechanics (Bourantas et al., 2018) but so far its effectiveness has been demonstrated only for linear problems. Element Free Galerkin (EFG) – based methods (Belytschko et al., 1994) appear the most attractive.

The Element Free Galerkin (EFG) method is an effective meshless method for nonlinear problems based on the diffuse elements method (DEM) originated by (Nayroles et al., 1992). The solution procedure of the EFG method is similar to that used in Finite Element Methods (FEM). However, in EFG the problem domain discretization is achieved using nodes arbitrarily distributed within and on the boundary of the problem domain. Galerkin weak form is employed to develop the discretized system of equations and background cells are used for numerical integration. The complex finite element grid generation and element distortion problem are eliminated, as no mesh for interpolating variable of interest (i.e. displacements) is required. The meshless approximation functions are constructed by using these arbitrarily distributed *field nodes*. Motivated by these prospects, (Horton et al., 2010b) developed the Meshless Total Lagrangian Explicit Dynamics (MTLED) algorithm based on the finite element TLED algorithm (Miller et al., 2007).

However, to reliably use meshless methods based on EFG formulation three long-standing challenges have to be met: i) meshless shape functions using higher order basis cannot always be computed for arbitrarily distributed nodes, i.e. not all node distributions are *admissible* (Gu, 2005; Joldes et al., 2015a). Yet, automatic and irregular node placement is crucial for patient-specific computational grid generation of complicated geometries from medical images; ii) difficulty in imposing the Essential Boundary Conditions (EBC), as the meshless shape functions are not interpolating; and iii) inexact numerical (Gauss) integration in space, as meshless shape functions are rationals (Gu, 2005; Li and Liu, 2004; Liu and Gu, 2005).

The Moving Least Squares (MLS) (Lancaster and Salkauskas, 1981; Shepard, 1968) has been the preferred choice of approximation in EFG due to its continuity and smoothness. However, the use of MLS with higher order polynomial basis (which offer higher accuracy) is not trivial for arbitrarily distributed nodes as many arbitrary generated clouds of points are found in practice *inadmissible* (Liu, 2003). To increase the proportion of *admissible* nodal distributions, we developed a Modified Moving Least Squares (MMLS) approximation (Joldes et al., 2015a).

The next limitation is the exact imposition of EBC which is crucial for accurate prediction of organ deformations during surgery. For some classes of problems (including the image registration where we are interested in deformation field within the organ rather than stresses and forces) driving deformation through EBC imposition can achieve accurate solutions without patient-specific information about the tissue constitutive properties (Miller and Lu, 2013; Wittek et al., 2009). Therefore, imposing EBC is crucial for patient-specific applications. However, the MLS and MMLS-derived shape functions are non-interpolating and do not possess Kronecker Delta property. Thus, imposing EBC in meshless methods is not as trivial as in the FEA. Most methods proposed for imposing EBC in meshless methods are not applicable to explicit time integration that enables real-time computations for surgery simulation on commodity hardware (off-the-shelf Graphics Processing Units) (Joldes et al., 2010b). To allow exact imposition of EBC for meshless methods using explicit time stepping, we have developed a new technique called Essential Boundary Conditions Imposition in Explicit Meshless (EBCIEM) method (Joldes et al., 2017).

Another difficulty is the numerical integration of rational shape functions emerging from MLS and MMLS, as well as other meshless approximants. Gauss quadratures are inexact for such functions. Therefore, spatial integration is a possible source of additional error, not normally present in finite element methods (isogeometric finite element analysis, however, faces the same difficulty (Cottrell et al., 2009). This difficulty was addressed by a new adaptive quadrature algorithm developed by (Joldes et al., 2015b).

The objectives of this paper are two-fold. We describe in detail the Meshless Total Lagrangian Explicit (MTLED) suite of numerical algorithms allowing accurate and reliable calculation of large deformations of soft tissues. We also demonstrate translational benefits of using our suite in patient-specific applications of clinical relevance.

The paper is organized as follows: the MTLED algorithms are presented in Section 2; numerical examples for verification against the Finite Element Method (ABAQUS) for moderate deformations are given in Section 3; experimental validation for very large deformations, where finite element method fails, are presented in Section 4; Section 5 contains an example of application of MTLED for biomechanics-based preoperative MRI to intraoperative CT neuroimage registration, a procedure of crucial importance for neuronavigation in epilepsy surgery. We discuss our results and present conclusions in Section 6.

2. **MTLED Suite of Algorithms**

*2.1 Modified Moving Least Squares Approximation*

The procedure for constructing Modified Moving Least Squares (MMLS) (Joldes et al., 2015a) shape function starts with the approximation of a function $u(\mathbf{x})$, denoted by $u^h(\mathbf{x})$, which is defined by a combination of *m* monomials, also known as basis functions:

$$u^h(\mathbf{x}) = \sum_{i=1}^{m} p_i(\mathbf{x})a_i(\mathbf{x}) = \mathbf{p}^T(\mathbf{x})\boldsymbol{a}(\mathbf{x}) \qquad (1)$$

with *m* being the number of terms in the basis $\mathbf{p}(\mathbf{x})$, and $a_i(\mathbf{x})$ coefficients that depend on the spatial coordinates **x**. These coefficients are computed by minimizing an error functional defined based on the weighted least squares errors and including additional constraints on the coefficients $\boldsymbol{a}$ corresponding to the second degree monomials in the basis. In 2D, the error functional is:

$$J(\mathbf{x}) = \sum_{j=1}^{n}\left[\left(u^h(\mathbf{x}_j) - u_j\right)^2 w(\|\mathbf{x} - \mathbf{x}_j\|)\right] + \mu_{x^2}a_{x^2}^2 + \mu_{xy}a_{xy}^2 + \mu_{y^2}a_{y^2}^2 \qquad (2)$$

where $n$ is the number of nodes in the support domain of $\mathbf{x}$ and $\boldsymbol{\mu} = [\mu_{x^2}\ \mu_{xy}\ \mu_{y^2}]$ are the positive weights for the additional constraints. For the 3D case, the error functional is defined as:

$$J(\mathbf{x}) = \sum_{j=1}^{n}\left[\left(u^h(\mathbf{x}_j) - u_j\right)^2 w(\|\mathbf{x} - \mathbf{x}_j\|)\right] + \mu_{x^2}a_{x^2}^2 + \mu_{y^2}a_{y^2}^2 + \mu_{z^2}a_{z^2}^2 + \mu_{xy}a_{xy}^2 + \mu_{xz}a_{xz}^2 + \mu_{yz}a_{yz}^2 \qquad (3)$$

with $\boldsymbol{\mu} = [\mu_{x^2}\ \mu_{y^2}\ \mu_{z^2}\ \mu_{xy}\ \mu_{xz}\ \mu_{yz}]$. After minimization and solving the resulting systems of equations, the MMLS approximation is obtained as:

$$u^h(\mathbf{x}) = \mathbf{p}^T(\mathbf{P}^T\mathbf{W}\mathbf{P} + \mathbf{H})^{-1}\mathbf{P}^T\mathbf{W}\mathbf{u} = \sum_{j=1}^{n}\emptyset_j(\mathbf{x})u_j = \boldsymbol{\Phi}^T(\mathbf{x})\mathbf{u} \qquad (4)$$

where $\mathbf{u}$ is the vector collecting the nodal parameters of the field variables for all the nodes in the local support domain and $\boldsymbol{\Phi}$ are the shape functions:

$$\boldsymbol{\Phi} = [\emptyset_1(\mathbf{x})\ \ldots\ \emptyset_n(\mathbf{x})] = \mathbf{p}^T(\mathbf{P}^T\mathbf{W}\mathbf{P} + \mathbf{H})^{-1}\mathbf{P}^T\mathbf{W} \qquad (5)$$

For 2D, $\mathbf{H}$ is a $6 \times 6$ matrix with all elements zeros except the last three diagonal entries, which are equal to the positive weights of the additional constraints $\boldsymbol{\mu}$:

$$H = \begin{bmatrix} \mathbf{0}_{33} & \mathbf{0}_{33} \\ \mathbf{0}_{33} & diag(\boldsymbol{\mu}) \end{bmatrix} \qquad (6)$$

For 3D, $\mathbf{H}$ is a $10 \times 10$ matrix with all elements zeros except the last six diagonal entries equal to $\boldsymbol{\mu}$:

$$H = \begin{bmatrix} \mathbf{0}_{44} & \mathbf{0}_{46} \\ \mathbf{0}_{64} & diag(\boldsymbol{\mu}) \end{bmatrix} \qquad (7)$$

The choice of weight function is more or less arbitrary as long as the weight function is positive and continuous together with its derivatives up to the desired order. We use quartic spline weight function in the construction of our MMLS approximation.

The choice of the additional constraints ensures that, when the classical MLS moment matrix is singular (multiple solutions), we obtain the solution having the coefficients for the higher order monomials in the bases equal to zero. By choosing the additional weights as small positive numbers we can ensure that the classical MLS solution is altered only very slightly when the moment matrix is not singular.

## 2.2 Total Lagrangian Explicit Dynamics

We use the Total Lagrangian (TL) formulation (Horton et al., 2010a; Miller et al., 2006) where all the calculations refer to the initial configuration of the analysed continuum. All derivatives with respect to spatial coordinates are computed during the pre-processing stage. This eliminates the necessity of such expensive computations at every time step as is the case when using the Updated Lagrangian formulation. After introducing MMLS approximation into the weak form of governing equations of solid mechanics using the TL formulation, the global system of discretized equations describing the behavior of the analyzed continuum becomes the following:

$$\mathbf{M}\,{}^t\ddot{\mathbf{u}} + {}^t\mathbf{F}_{int} = {}^t\mathbf{F}_{ext} \qquad (8)$$

where $\mathbf{u}$ is the vector of nodal displacements, $\mathbf{M}$ is the mass matrix, ${}^t\mathbf{F}_{int}$ is the global nodal reaction force vector and ${}^t\mathbf{F}_{ext}$ is the vector of externally applied force at time $t$. The vector of internal nodal forces (${}^t\mathbf{F}_{int}$) is computed as:

$$ {}^t\mathbf{F}_{int} = \int_{V_0} {}^t_0\mathbf{X}\,{}^t_0\mathbf{B}_{L0}^{\mathrm{T}}\,{}^t_0\mathbf{S}\,\mathrm{d}V_0 \qquad (9)$$

where ${}^t_0\mathbf{X}$ is the deformation gradient at time $t$, ${}^t_0\mathbf{S}$ is the second Piola-Kirchoff stress at time $t$, ${}^t_0\mathbf{B}_{L0}$ is the matrix of shape function derivatives and $V_0$ is the initial volume of the problem domain.

We apply explicit integration in time domain using central difference method. Explicit time integration is a direct integration method where nodal accelerations are found directly without any iteration and then integrated to obtain the displacements. This eliminates the need for assembling a global stiffness matrix. The time stepping scheme for solving the equation of motion can be expressed as:

$$^{t+1}\mathbf{u} = \Delta t^2 \mathbf{M}^{-1}(^{t}\mathbf{F}_{ext} - {}^{t}\mathbf{F}_{int}) + 2{}^{t}\mathbf{u} - {}^{t-1}\mathbf{u} \qquad (10)$$

where $^{t}\mathbf{u}$ is the displacement calculated at time $t$, $\mathbf{M}$ is the constant diagonal mass matrix and $\Delta t$ is the time step.

*2.3    Dynamic Relaxation*

Dynamic relaxation is an explicit iterative method for obtaining steady state solution, for a discretized continuum mechanics problem. In the Dynamic Relaxation (DR) algorithm (Joldes et al., 2009b, 2011), we introduce a damping force to the equation of motion to dissipate the kinetic energy when the steady state of the deformed continuum needs to be obtained. This is done by introducing mass proportional damping to enable the decoupling of equations for explicit time integration and efficient convergence to the steady state solution.

$$\mathbf{M}^{t}\ddot{\mathbf{u}} + c\mathbf{M}^{t}\dot{\mathbf{u}} = {}^{t}\mathbf{F}_{ext} - {}^{t}\mathbf{F}_{int} \qquad (11)$$

where $c\mathbf{M}^{t}\dot{\mathbf{u}}$ is the damping force and $c$ is the damping coefficient. The resulting equation describing the iterations in terms of displacements is derived as:

$$^{t+1}\mathbf{u} = {}^{t}\mathbf{u} + \beta({}^{t}\mathbf{u} - {}^{t-1}\mathbf{u}) + \alpha \mathbf{M}^{-1}({}^{t}\mathbf{F}_{ext} - {}^{t}\mathbf{F}_{int}) \qquad (12)$$

where $\alpha = 2h^2/(2 + ch)$, $\beta = (2 - ch)/(2 + ch)$ and $h$ is a fixed time increment.

In the relaxation stage, the integration time step $\Delta t$ is kept constant, while the damping coefficient $c$ and lumped mass matrix $\mathbf{M}$ are initiated following (Joldes et al., 2017) and automatically adjusted to maximize the convergence rate and improve the computational efficiency without compromising the solution convergence.

*2.4    Essential Boundary Conditions Imposition for Explicit MTLED*

In meshless methods, as in FEM, the imposition of natural boundary conditions does not present a problem. The difficulty in imposing Essential Boundary Conditions (EBC) in meshless methods arises from the properties of the meshless shape functions. In MTLED (and most other meshless methods), the shape functions are created with overlapping support domains and are generally not interpolating at nodes. Thus, prescribing certain values of the field variables will not yield the exact nodal displacements on the boundary nodes. Furthermore, traditional methods (such as Lagrange multipliers (Liu, 2003, Belytschko et al., 1994) of imposing EBC are not applicable in explicit time integration framework.

We have introduced two new ways of imposing essential boundary condition in meshless method based on Element Free Galerkin principle and suitable for explicit time integration framework (Joldes et al., 2016). The new methods (referred to as *Essential Boundary Conditions Imposition for Explicit Meshless* and *Simplified Essential Boundary Conditions Imposition for Explicit Meshless*) consider the external forces on the essential boundary as additional unknowns, which are later eliminated from the time-discretized equation of motion using static condensation (Bathe, 1996). Therefore, we split the total externally applied force in the equation of motion into two parts ${}^t\mathbf{F}_{ne} + {}^t\mathbf{F}_e = {}^t\mathbf{F}_{ext}$. The equation of motion can then be express as:

$$\begin{aligned} \mathbf{M}{}^t\ddot{\mathbf{u}} + c\mathbf{M}{}^t\dot{\mathbf{u}} &= ({}^t\mathbf{F}_{ne} - {}^t\mathbf{F}_{int}) + {}^t\mathbf{F}_e \\ {}^t\mathbf{u} &= {}^t\bar{\mathbf{u}} \; on \; \Gamma_e \end{aligned} \qquad (13)$$

where ${}^t\mathbf{F}_e$ is the force that is externally applied only on the essential boundary $\Gamma_e$, ${}^t\mathbf{F}_{ne}$ is the force that is applied on the non-essential boundary, and ${}^t\bar{\mathbf{u}}$ is the value of the displacements on the essential boundary at time t. In EFG method, ${}^t\mathbf{F}_e$ is calculated as:

$${}^t\mathbf{F}_e = \int_{\Gamma_e} {}^t\boldsymbol{\Phi}(s) \cdot {}^t\boldsymbol{T}(s) \cdot d\Gamma_e \qquad (14)$$

where ${}^t\boldsymbol{\Phi}(s)$ are the meshless shape functions, $s$ is the arc-length along the essential boundary, $T$ is the distributed force on the essential boundary. We express the externally applied force on the essential boundary (${}^t\mathbf{F}_e$) using two methods.

In the *Essential Boundary Conditions Imposition for Explicit Meshless* (EBCIEM), the externally applied forces are considered as distributed force and values of the distributed force are interpolated at the essential boundary nodes. A discretization along the essential boundary is necessary in this case to numerically integrate the externally applied force on the essential boundary. In this case, the externally applied force on the essential boundary is obtained as:

$$^t\mathbf{F}_e = \sum_{i=1}^{n_g} {}^t\boldsymbol{\Phi}(s_i) \cdot \sum_{k=1}^{n_e} N^k(s_i) \cdot {}^t\mathbf{T}_e^k \cdot w_i \qquad (15)$$

where $N(s)$ are the shape functions used for interpolation, $n_e$ is the number of essential boundary nodes, $s_i$ and $w_i$ are the Gauss quadrature points and weights respectively and $n_g$ is the total number of integration points along the essential boundary segment. Eq. (15) is written in matrix form as:

$$^t\mathbf{F}_e = {}^t\mathbf{V} \cdot {}^t\mathbf{T}_e \qquad (16)$$

with

$$^t\mathbf{V}_{jk} = \sum_{i=1}^{n_g} {}^t\boldsymbol{\Phi}_j(s_i) \cdot N^k(s_i) \cdot w_i \ . \qquad (17)$$

In the *Simplified Essential Boundary Conditions Imposition for Explicit Meshless* (SEBCIEM), the distributed forces on the essential boundary are lumped at the essential boundary nodes. The advantage of SEBCIEM is that it does not require any discretization along the essential boundary to evaluate the externally applied force. In this case, the externally applied force on the essential boundary is obtained as:

$$^t\mathbf{F}_e = \sum_{k=1}^{n_e} {}^t\boldsymbol{\Phi}(s_k) \cdot \mathbf{T}_e^k. \qquad (18)$$

Eq. (18) is written in matrix form as:

$$^t\mathbf{F}_e = {}^t\mathbf{V} \cdot {}^t\mathbf{T}_e \qquad (19)$$

with

$$^t\mathbf{V}_{jk} = {}^t\boldsymbol{\Phi}_j(s_k). \qquad (20)$$

To obtain the equation describing the iterations in terms of displacements, the unknown forces on the essential boundary are eliminated from the system of equations describing the time discretization of the equation of motion (using the central difference method), and augmented with the equations defining the imposed displacements on the boundary, Eq 21, 22:

$$^{t+1}\mathbf{u} = \;^{t+1}\widetilde{\mathbf{u}} + \mathbf{M}^{-1}.\,^{t}\mathbf{V}\,[\,\boldsymbol{\Phi}\,.\,\mathbf{M}^{-1}.\,^{t}\mathbf{V}\,]^{-1}[\,^{t+1}\overline{\mathbf{u}} - \;\boldsymbol{\Phi}\,.\,^{t+1}\widetilde{\mathbf{u}}] \qquad (21)$$

with

$$^{t+1}\widetilde{\mathbf{u}} = \;^{t}\mathbf{u} + \beta(^{t}\mathbf{u} - \;^{t-1}\mathbf{u}) + \alpha\mathbf{M}^{-1}.\,(^{t}\mathbf{F}_{ne} - \;^{t}\mathbf{F}_{int}) \qquad (22)$$

where $\widetilde{\mathbf{u}}$ is the predicted displacement when the load on the essential boundary is disregarded. Eq (21) is rewritten as:

$$^{t+1}\mathbf{u} = \;^{t+1}\widetilde{\mathbf{u}} + \;^{t+1}\mathbf{u}_{correction} \qquad (23)$$

where

$$^{t+1}\mathbf{u}_{correction} = \;^{t}\mathbf{P}.\,[\,^{t+1}\overline{\mathbf{u}} - \;\boldsymbol{\Phi}\,.\,^{t+1}\widetilde{\mathbf{u}}] \qquad (24)$$

with

$$^{t}\mathbf{P} = \;\mathbf{M}^{-1}.\,^{t}\mathbf{V}\,[\,\boldsymbol{\Phi}\,.\,\mathbf{M}^{-1}.\,^{t}\mathbf{V}\,]^{-1} \qquad (25)$$

In the context of TL settings, $^{t}\mathbf{P}$ is a constant matrix which can be precomputed because the MMLS shape functions do not change during time-stepping. Both EBCIEM and SEBCIEM define displacement corrections which are added to the displacement field during time-stepping.

## 2.5    *Adaptive Spatial Integration*

In FEM, the integration cells coincide with the element mesh and shape functions are polynomials over the integration cells, therefore the application of Gauss quadratures to yield exact integration results is straightforward. In MTLED (and most other meshless methods), Gaussian quadrature over a background mesh (not needing to meet criteria of quality as finite element meshes do) is used for numerical integration. Unlike the FEM shape functions, MLS and MMLS shape functions are not polynomials but rationals and they usually have a much larger support domain which may not align with the integration cells. These may lead to integration inaccuracies in EFG based meshless methods.

MTLED uses a new adaptive quadrature algorithm for EFG methods (Joldes et al., 2015b). The algorithm creates a distribution of integration points within the problem domain and allows the computation of integrals with controlled accuracy. The method introduces new integration points only in the areas where the integration accuracy is not sufficient (does not satisfy the required accuracy threshold). The method imposes no constraints on the type of support domains that can be used.

In the TL formulation, numerical integration is required to evaluate the global nodal reaction force vector, which is defined as:

$$ {}_0^t\mathbf{F} = \int_{V_0} {}_0^t\mathbf{B}_L^T \, {}_0^t\mathbf{S} \, dV_0 \tag{26}$$

with,

$$ {}_0^t\mathbf{B}_L = [{}_0^t\mathbf{B}_L^{(1)}, {}_0^t\mathbf{B}_L^{(2)}, \ldots, {}_0^t\mathbf{B}_L^{(n)}] \tag{27}$$

$$ {}_0^t\mathbf{B}_L^{(i)} = {}_0^t\mathbf{B}_{L0}^{(i)} \, {}_0^t\mathbf{X}^T \tag{28}$$

where ${}_0^t\mathbf{B}_L$ is the full strain-displacement matrix, ${}_0^t\mathbf{S}$ is the second Piola–Kirchoff stress vector and ${}_0^t\mathbf{X}$ is the deformation gradient. ${}_0^t\mathbf{B}_{L0}$ is the matrix of shape function derivatives in reference to the initial configuration and has the following form:

$$ {}_0^t\mathbf{B}_{L0}^{(i)} = \begin{bmatrix} \emptyset_{,x} & 0 & 0 \\ 0 & \emptyset_{,y} & 0 \\ 0 & 0 & \emptyset_{,z} \\ \emptyset_{,y} & \emptyset_{,x} & 0 \\ 0 & \emptyset_{,z} & \emptyset_{,y} \\ \emptyset_{,z} & 0 & \emptyset_{,x} \end{bmatrix} \tag{29}$$

where $\emptyset_{,x}$, $\emptyset_{,y}$ and $\emptyset_{,z}$ are derivatives of shape functions with respect to x, y and z respectively and their values are taken directly from the precomputed MMLS shape function derivatives matrix $D\Phi(\mathbf{x})$.

In the adaptive quadrature method (Joldes et al., 2015b), a function of 'less smooth' integrand needs to be defined. This function is integrated to a user-defined accuracy. This generates a collection of integration points and weights over the integration cell. This 'less smooth' idea is based on the observation that if the adaptive integration procedure is able to accurately integrate a given integrand over the integration cell, it should also accurately integrate integrands that are 'smoother'. For example, as discussed in (Joldes et al., 2015b), if function *f* and *g* are approximated by the polynomial of degree *n* and *m* respectively, then their product *fg* can be approximated by a polynomial *p* of degree *m+n*. At least one of $f^2$ and $g^2$ are less smooth than *fg* as they require for approximation, a polynomial of degree higher than the degree of polynomial *p*. Moreover, the degree of a polynomial approximation for $f^2+g^2$ is *max(2m,2n)*. Following the observations and considering Eq(26)-(29), we define this 'less smooth' integrand function based on the MMLS shape function derivatives as:

$$f(x) = \begin{bmatrix} \sum_{k=1}^{n} \left(\phi^k_{,x}(\mathbf{x})\right)^2 \\ \sum_{k=1}^{n} \left(\phi^k_{,y}(\mathbf{x})\right)^2 \\ \sum_{k=1}^{n} \left(\phi^k_{,z}(\mathbf{x})\right)^2 \end{bmatrix} \qquad (30)$$

After defining the less smooth integrand, we apply the procedure for performing the adaptive numerical integration in MTLED by the following algorithm:

*Step 1:* Pre-computation of integration points (*xi*) and associated weights (*wi*):
- Select desired accuracy ($\tau$)
- Select a quadrature rule to apply to the integration cells.
- For each initial integration cell, apply the recursive integration procedure for the function *f* in Eq(30) used to drive the subdivisions:
    o Select a scheme to geometrically subdivide the initial integration cell (*cell*) into some number of subdivisions (*cell₁, cell₂…cellₙ*).
    o Approximate the integral in the initial cell by the selected quadrature rule.
        [*Q, xi, wi*] = integrate (*f, cell*)
    o Approximate the integrals in the subdivided cells.
        [$Q_1$, $xi_1$, $wi_1$] = integrate (*f, cell₁*)
        [$Q_2$, $xi_2$, $wi_2$] = integrate (*f, cell₂*)
        …
        [$Q_n$, $xi_n$, $wi_n$] = integrate (*f, cellₙ*)
    o Find the error, $\varepsilon = \left|\frac{Q-\sum_{i=1}^{n} Q_i}{Q}\right|$
    o If ( $\varepsilon > \tau$ ), return:

- $Q = \sum_{i=1}^{n} Q_i$
- $xi$ = concatenate ( $xi_1$ , $xi_2$ , … , $xi_n$ )
- $wi$ = concatenate ( $wi_1$ , $wi_2$ , … , $wi_n$ )

*Step 2:* Perform numerical integration
- Compute the global nodal reaction force vector over the problem domain by the integration points ($x_i$) and weights ($w_i$) determined in Step 1 as:

$$ {}_{0}^{t}\mathbf{F} = \sum_{i=1}^{N_{ip}} ( {}_{0}^{t}\mathbf{X} \, {}_{0}^{t}\mathbf{B}_{L0}^{\mathrm{T}} \, {}_{0}^{t}\mathbf{S} )|_{x=xi} \, wi \qquad (31) $$

where $N_{ip}$ is the total number of integration points distributed in the problem domain. In the TL settings, there is no necessity to update the distribution of integration points as the MMLS shape functions and their derivatives do not change during the solution process. All elaborate calculations given in above are conducted only once, in the pre-processing stage.

## 3 Verification and Validation of MTLED

In this section, numerical examples are presented verifying MTLED against FEM for moderate deformations, where FEM gives reliable results. We consider i) unconstrained compression of a cube in 3D; ii) extension and compression of a sheep brain sample in 3D; iii); swine brain indentation in 3D.

In our adaptive integration procedure, for a given number of recursive levels, the initial integration domain is subdivided into a number of subdomains at each level. For the 3D cases the tetrahedral background integration cells (that DO NOT need to have shape quality required by FEM) are used. During the adaptive procedure, the integration cells are subdivided into either 2 or 4 or 8 subdomains. For the case where the original integration cell is subdivided into 2 subdomains, a new point (vertex corner) is introduced at the midpoint of the longest edge of the original tetrahedral cell. For 4 subdivisions, 3 new points are created at the midpoints of the three edges of the original tetrahedral cell. Similarly, midpoints of all the edges of the original tetrahedral integration cell are connected to create 8 subdivisions within the mother-cell.

In section 3.2 and 3.3, the meshless results are compared with the results obtained using ABAQUS (version 16.14-1) (ABAQUS, 2018) - commercial FE software considered reliable in non-linear analysis and widely used in computational biomechanics (Miller and Nielsen, 2010). Identical geometry, material models and material properties are used in meshless and FE computations. In the FE computations, we used a non-linear static procedure with direct solver and Newton's iterative method available in ABAQUS.

*3.1 Unconstrained Compression of a Cube*

In this example, an unconstrained compression of a cube (edge length of 100 mm) having brain-tissue-like constitutive properties is modelled. Analytical solution for vertical component of displacement is available for this simple case. The meshless discretizations of the problem domain (260 and 3,364 nodes) and the boundary conditions are as shown in Figure 1. We used MMLS approximation method with a constant influence domain for all nodes and the same weights for all the additional MMLS constraints ($\mu=10^{-7}$).

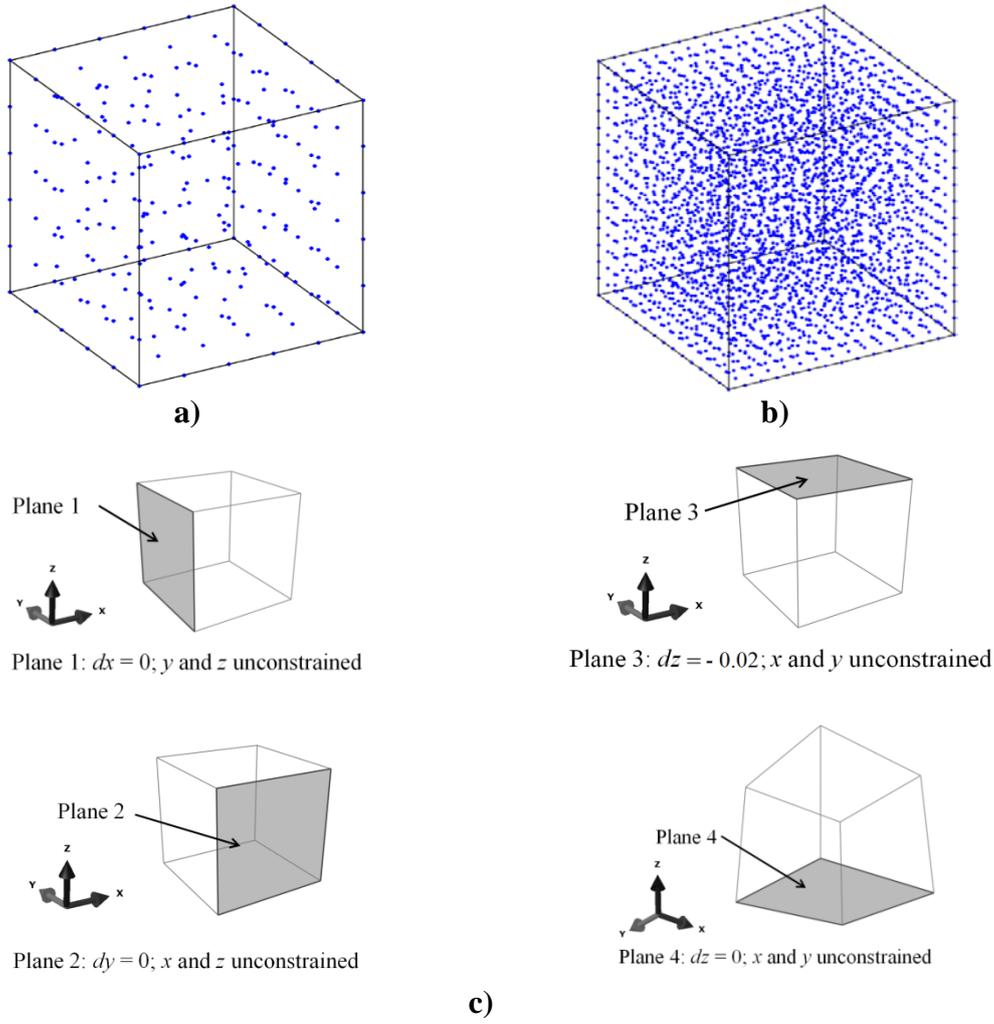

**Figure 1**: The computational grid: **a)** 260 nodes; **b)** 3,364 nodes; and **c)** the boundary conditions for modelling of unconstrained compression of a cube with brain-tissue-like constitutive properties.

The EBCIEM method is used to impose the Essential Boundary Conditions (EBC). The constitutive behavior is described using a hyper-elastic Neo-Hookean material model with Young′s modulus of 3000 Pa, Poisson's ratio of 0.49 and density of 1000 kg/m³. The cube is compressed by displacing the top surface by 20 mm (i.e. 20% of the initial height). We use the Normalized Root Mean Square Error $NRMSE = \dfrac{\sqrt{\frac{1}{N}\sum_{i=1}^{N}\left(u_i^{MTLED}-u_i^{analytical}\right)}}{u_{max}^{analytical}-u_{min}^{analytical}}$ as a measure of accuracy. Table 1 shows the NRMSE for different choices of the number of nodes and integration points.

**Table 1**: NRMSE error norm between MTLED (models from Figure 1) and analytical solution (z- direction).

| Approximation method | Basis function | Integration points | NRMSE $u_z$ |
|---|---|---|---|
| MMLS | Quadratic | 1,014 | $3{,}08 \times 10^{-3}$ |
|  |  | 4,056 (tetrahedral integration cells spun over **260 nodes (Fig. 1a)**, no subdivision) | $6{,}92 \times 10^{-4}$ |
| MMLS | Quadratic | 66,112 (tetrahedral integration cells spun over **3,364 nodes (Fig 1b)**, no subdivision) | $2{,}94 \times 10^{-4}$ |

**Adaptive integration (260 nodes)**

| Approximation method | Accuracy | Integration points | NRMSE |
|---|---|---|---|
| MMLS (Quadratic) | 0.1 | 4,992 | $6{,}30 \times 10^{-4}$ |
|  | 0.05 | 16,284 | $3{,}45 \times 10^{-4}$ |
|  | 0.01 | 71,352 | $1{,}04 \times 10^{-4}$ |

The results indicate that even using a very coarse grid of only 260 points with 1,014 single integration point tetrahedral used for integration give acceptable accuracy. Specifying high spatial integration accuracy increases the overall precision of the algorithm, however for applications in image-guided surgery, where the typical voxel sizes are of the order of 1 mm³, even very coarse grids with few integration points may be sufficient.

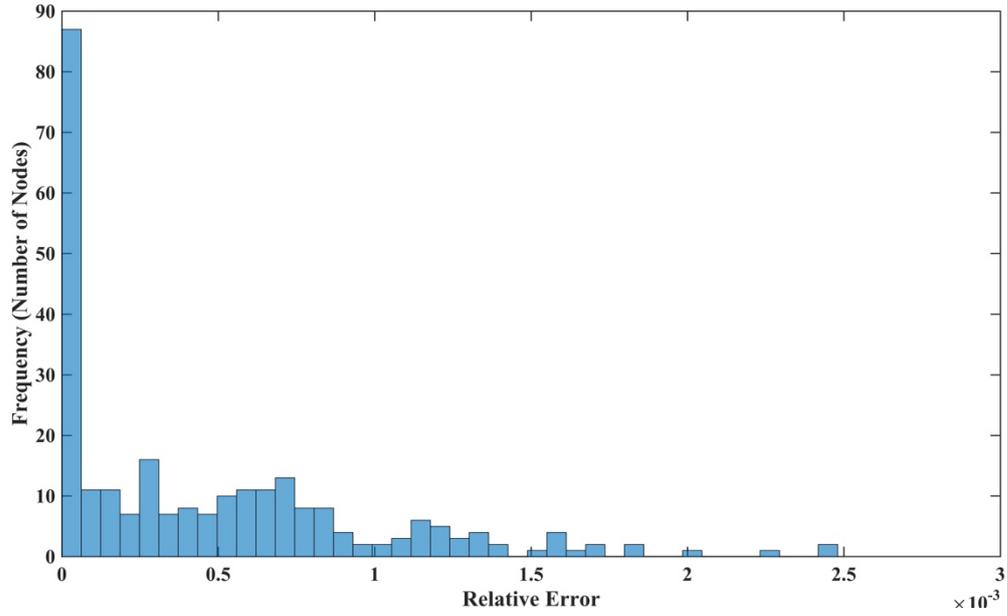

a)

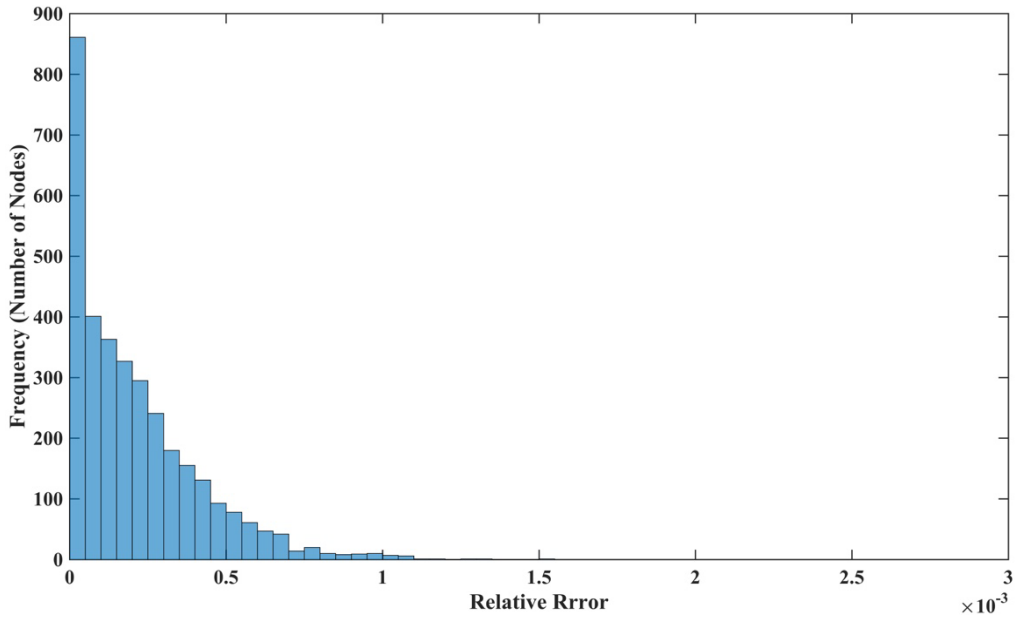

b)

**Figure 2**: **a)** A histogram displaying the normalized relative errors node by node for a model with 260 nodes and **b)** with 3,364 nodes for deformation field in the z-direction (the coordinate system is in Figure 1).

Moreover, as seen in Figure 2, the accuracy of the solution for most of the nodes is higher than that suggested by the normalized relative error ($NRE = \left|\frac{u_i^{MTLED} - u_i^{analytical}}{u_{max}^{analytical} - u_{min}^{analytical}}\right|$).

It is also worth noting, that our method of essential boundary condition imposition is exact, as the displacement error of nodes on the bottom and top surfaces of the cube is zero.

*3.2 Extension and Compression of a Sheep Brain Sample in 3D*

In this example, the extension of a brain tissue sample (sheep brain) is modelled (Agrawal et al., 2015). The sheep head specimen was collected from Royal Perth Hospital, as a by-product of anesthesia training program. The specimen was not frozen at any time. The head was skinned and a rectangular cut was made on the skull on top of the cranium using a vibrating saw. Then using a microtome blade the underlying brain tissue was extracted through the opening of the skull and the top surface of the brain was carefully levelled. Only the dimension of the brain tissue is taken into account excluding the skull and meninges (the brain tissue sample is modelled and the skull and meninges are disregarded). The sample geometry (undeformed configuration) is shown in Figure 3a and the undeformed and deformed configuration of the sample model in Figures 3b and 3c.

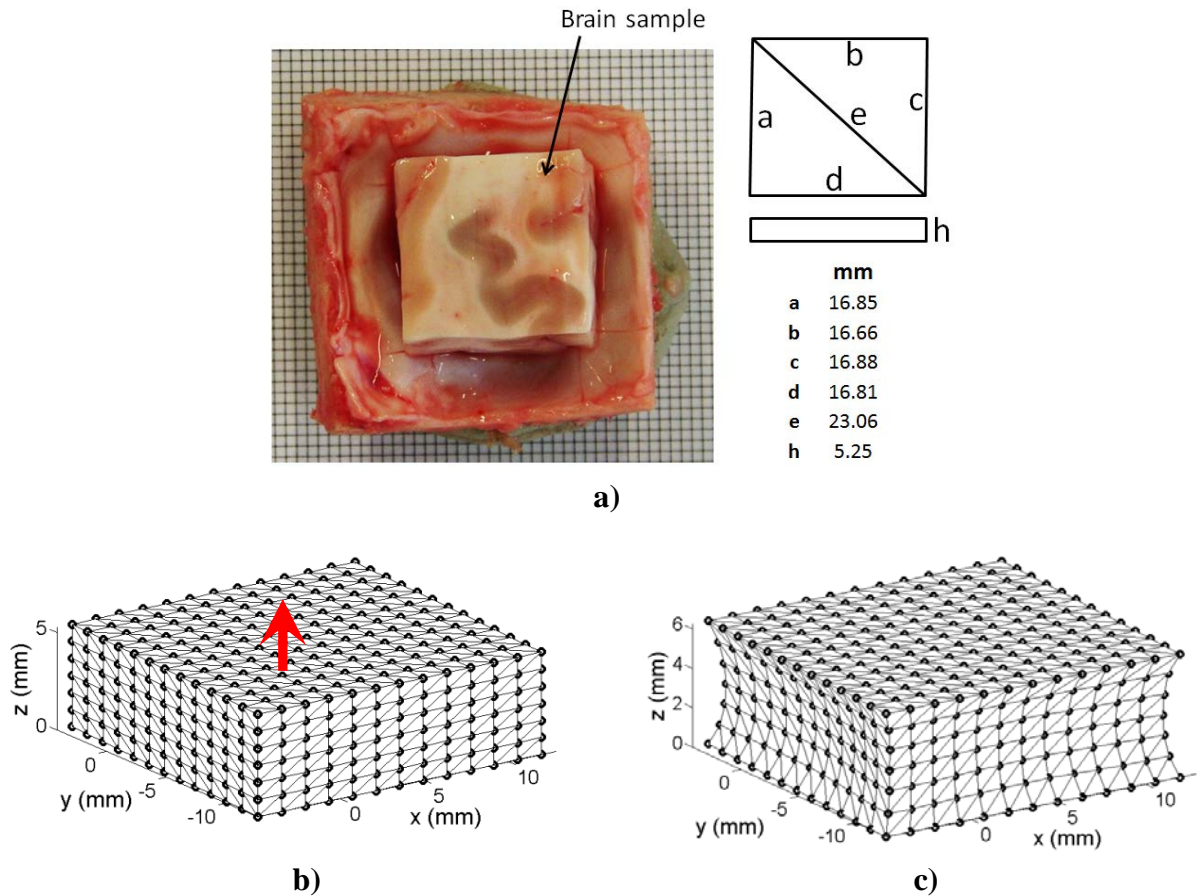

**Figure 3**: Extension of a sheep brain sample: **a)** Dimensions of the sheep brain sample, **b)** model in undeformed configuration, and, **c)** model in deformed configuration.

The base (z = 0 mm) of the brain tissue sample is rigidly constrained while the top surface (z = 5.25 mm) is displaced. Extension loading (1 mm) is applied smoothly, using a 3-4-5 polynomial (Waldron and Kinzel, 2004). The hyper-elastic Neo-Hookean material model with Young′s modulus of 3000 Pa, Poisson's ratio of 0.49 and density of 1000 kg/m$^3$ is chosen. We used two meshless discretizations: i) 1,085 nodes and 18,556 integration points; and ii) 13,073 nodes and 264,856 integration points. MMLS shape functions with a constant influence domain (R= 0.0022) were used. EBCIEM was used to enforce the essential boundary conditions. For verification, the meshless results are compared with the solution obtained using non-linear finite element model consisting of 66,214 hybrid tetrahedral elements (C3D4H in ABAQUS).

**Table 2**: Extension of a Sheep Brain Sample in 3D - Differences in nodal displacements between the meshless (MTLED) and ABAQUS solutions.

| Approximation method | Basis function | Integration points | NRMSE | | |
|---|---|---|---|---|---|
| | | | $u_x$ | $u_y$ | $u_z$ |
| MMLS | Quadratic | 18,556 (tetrahedral integration cells spun over 1085 nodes) | $8{,}44 \times 10^{-3}$ | $8{,}76 \times 10^{-3}$ | $2{,}53 \times 10^{-2}$ |
| | | 264,856 (tetrahedral integration cells spun over 13,073 nodes) | $1{,}75 \times 10^{-3}$ | $1{,}70 \times 10^{-3}$ | $6{,}99 \times 10^{-3}$ |

Table 2 shows the differences (summarized as $= \frac{\sqrt{\frac{1}{N}\sum_{i=1}^{N}\left(u_i^{MTLED}-u_i^{ABAQUS}\right)}}{u_{max}^{ABAQUS}-u_{min}^{ABAQUS}}$ ) of the computed deformation field between MTLED and FEM results.

Figure 4 visualizes the spatial distribution of the difference between MTLED and FEM results (normalized relative error ($NRE = \frac{u_i^{MTLED}-u_i^{ABAQUS}}{u_{max}^{ABAQUS}-u_{min}^{ABAQUS}}$). It is clear that near singularities at the edges the errors are larger and that far from singularities the MTLED solution is very accurate. It appears that, for most practical purposes, accuracy obtained with a coarse grid of 1,085 nodes should be acceptable.

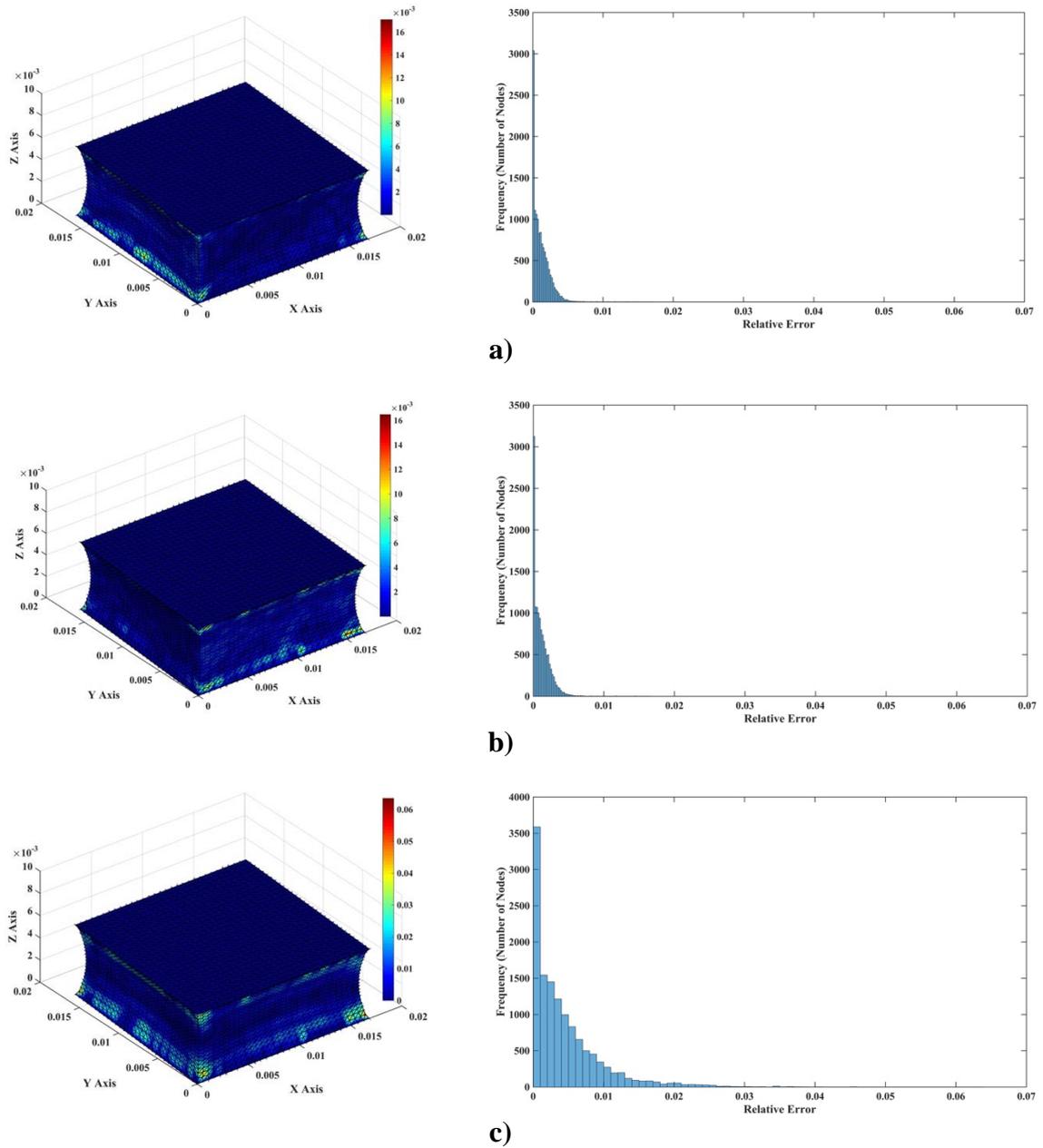

**Figure 4**: Differences (measured using normalized relative error) of the computed deformation field between MTLED (13,073 nodes and 264,856 integration points) and ABAQUS (66,214 hybrid tetrahedral elements C3D4H) (left-hand-side column) and histograms of the differences (right-hand-side column). **a)** *x* axis direction; **b)** *y* axis direction; **c)** z axis direction.

The results of modeling of compression of the brain tissue sample (13,073 nodes for both finite element and meshless models) are shown in Figure 5, Figure 6 and Table 3. The largest differences between the deformations obtained using MTLED and ABAQUS (hybrid tetrahedral elements C3D4H) can be observed near the top and bottom edges (see fringe plots of normalized relative error the right-hand-side of Figure 5). This is due to distortion of the of the hybrid tetrahedral elements along the edges (Figure 6). In MTLED such distortion of the computational grid was largely limited to the model corners (which, from numerical perspective, are singularities). Additionally, we present in Figure 7 the visualization of element pressure stress (negative one third trace of the stress tensor) when attempting this simulation with standard 4-noded linear tetrahedras – the element type frequently used in surgical simulation literature. The result is unphysical as can be seen from pressure jumping between positive and negative at neighboring elements. This is one of the instances of incorrect solution due to volumetric locking (Bathe, 1996).

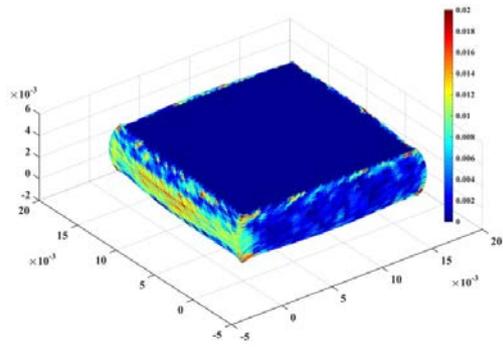 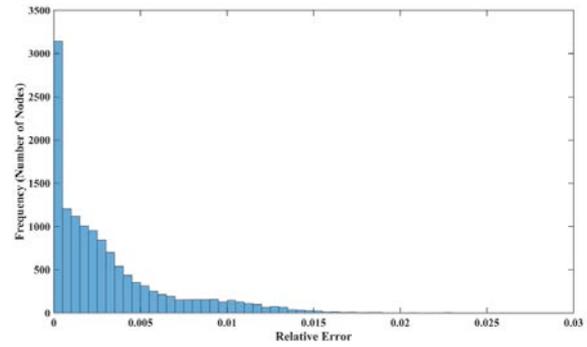

**a)**

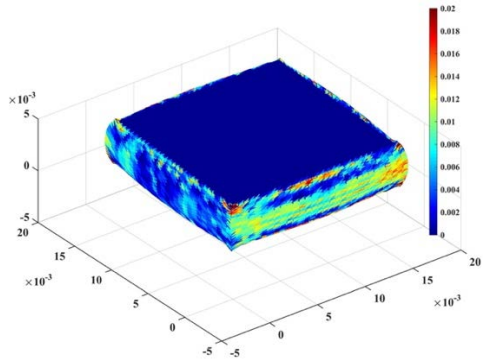 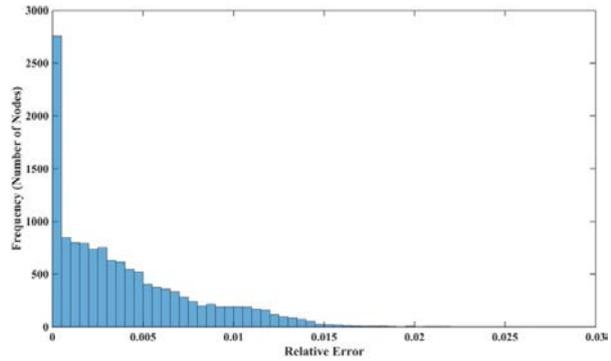

**b)**

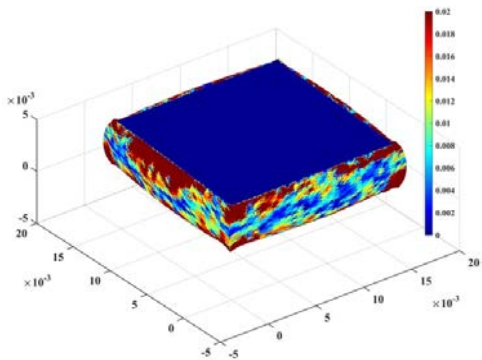 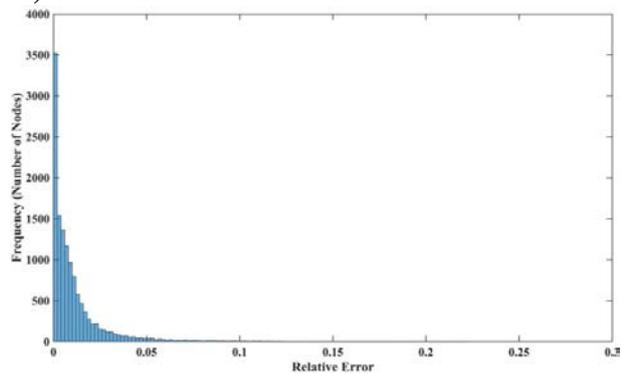

**c)**

**Figure 5**: Differences (measured using normalized relative error) of the computed deformation field between MTLED and ABAQUS (hybrid tetrahedral elements C3D4H) (left-hand-side column) and histograms of the differences (right-hand-side column) for **a)** x axis direction; **b)** y axis direction; **c)** z axis direction.

**Table 3:** Compression of a sheep brain sample - Differences in nodal displacements between the meshless (MTLED) and ABAQUS (four-noded hybrid tetrahedral elements) solutions.

| Approximation method | Basis function | Integration points | NRMSE | | |
|---|---|---|---|---|---|
| | | | $u_x$ | $u_y$ | $u_z$ |
| MMLS | Quadratic | 66,214 (tetrahedral integration cells spun over 13,073 nodes) | $7{,}74 \times 10^{-3}$ | $5{,}57 \times 10^{-3}$ | $2{,}03 \times 10^{-2}$ |

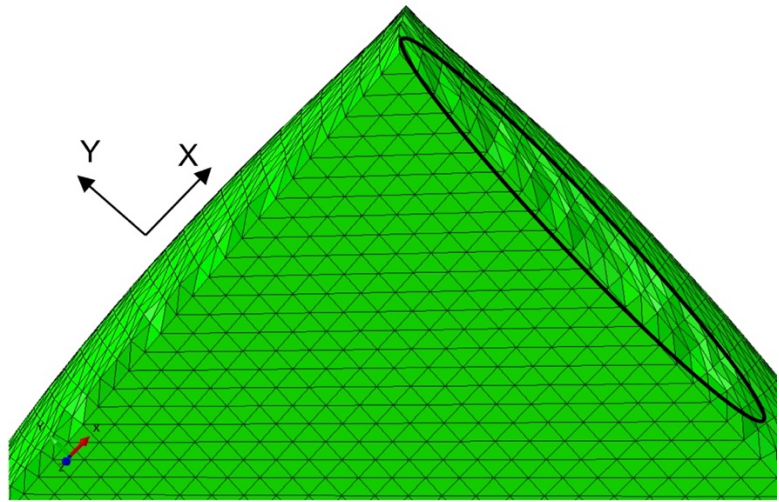

a)

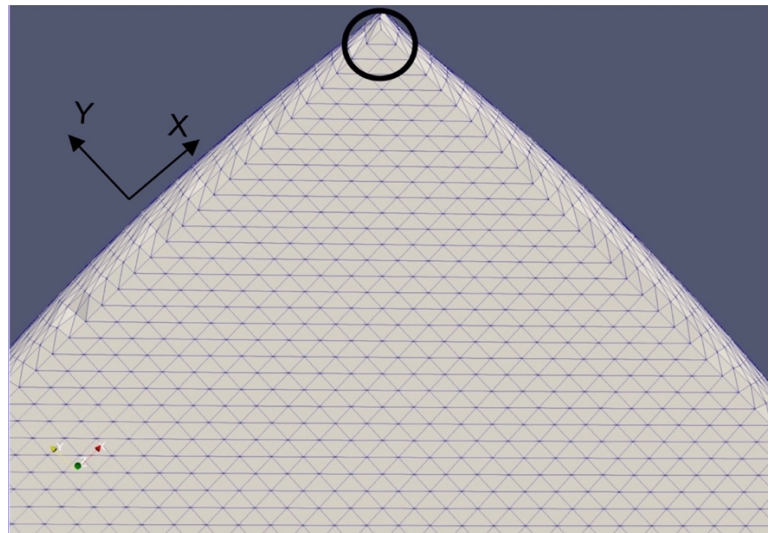

b)

**Figure 6**: Distortion of the computational grids when simulating compression of the brain tissue sample (bottom view). **a)** finite element (ABAQUS, hybrid tetrahedral elements) and **b)** MTLED (the field nodes were connected to form the triangles to visualize the deformed model surface, we used Paraview by Kitware). The areas of large distortion are indicated by the black ellipsoid and circle. The figure shows the results for compression of 20% of the sample height. It is clear that computational grid distortions close to the sample edge are much more severe for FEM than for MTLED.

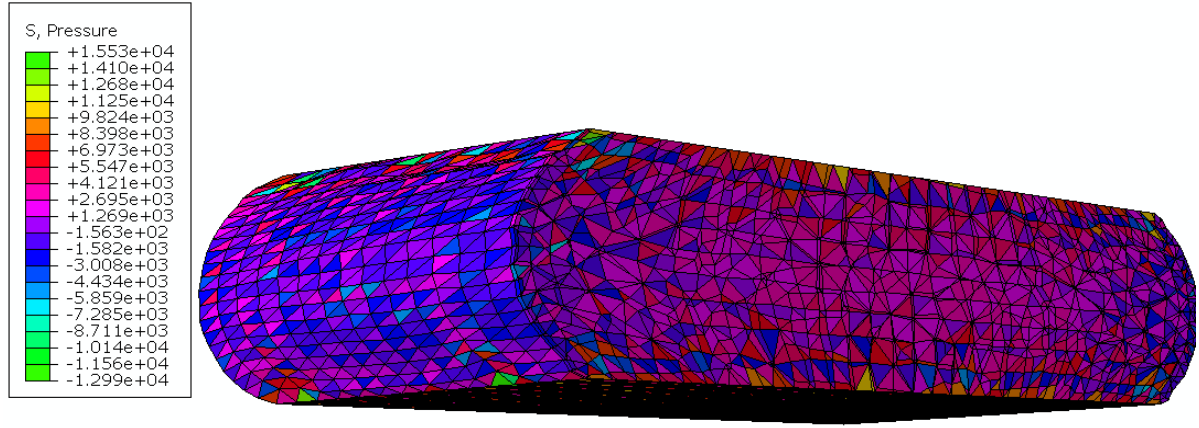

**Figure 7**: Results of simulation (non-linear static finite element analysis from ABAQUS code) of the brain tissue sample compression using 4-noded linear tetrahedral elements (element type C3D4 in ABAQUS code). The figure shows deformation and "pressure stress" (negative one third trace of the stress tensor in Pa) for the compression of 20% of the sample height. The checkerboard pressure distribution with the negative and positive pressure in adjacent elements is clearly visible. Such unphysical pressure distribution is a well-known indication of volumetric locking (artificial stiffening for incompressible/nearly incompressible materials) (Bathe, 1996). In the example shown in this figure, we used neo-Hookean constitutive model with the shear modulus $\mu$=1003.3 Pa and Poisson's ratio of 0.495. (This Figure should be viewed in color).

### 3.3 Swine Brain Indentation

In this experiment we demonstrate the usefulness of our methods for surgical simulation. We apply MTLED to model the previously conducted indentation experiments of the swine brain (Wittek et al., 2008a) and compare the results with the non-linear static finite element solver from ABAQUS.

The geometry of the brain was obtained from the MRIs as described in (Wittek et al., 2008a). In the meshless discretization we used 21,498 nodes and 115,029 tetrahedral background integration cells (Figure 8). For FEM solution we used 10-noded quadratic tetrahedral elements with hybrid (constant pressure) formulation — C3D10H element type in ABAQUS software. To facilitate comparisons, the vertices of finite elements coincided with the positions of nodes in MTLED model. The finite element mesh had 162,474 nodes.

In the experiments, the swine brain was constrained on its base by glue and a custom-made mold which is significantly stiffer than the brain tissue. To simulate a fixed base, all nodes on the bottom surface of the brain and the areas in contact with the mold are rigidly constrained. As our goal is evaluation of the performance of our meshless algorithms rather than modeling the interactions between the indenter and the brain, we prescribed the displacements (maximum of 5 mm, imposed smoothly using a 3-4-5 polynomial) on the nodes that were in contact with the indenter in the experiments instead of directly modelling the indenter, Figure 8. Following (Wittek et al., 2008a), the Neo-Hookean hyperelastic material model (shear modulus of 210 Pa and mass density of 1000 kg/m$^3$) was used for the brain tissue.

For the MTLED framework, we used the MMLS shape functions with quartic spline weight function. When prescribing the essential boundary conditions, we used the EBCIEM method that enforces such conditions exactly. The tetrahedral background integration cells and the adaptive integration procedure with the desired integration accuracy of 0.1% were used for spatial integration. This resulted in 528,152 integration points.

The difference in computed deformation field between meshless and ABAQUS results are shown in Figure 9.

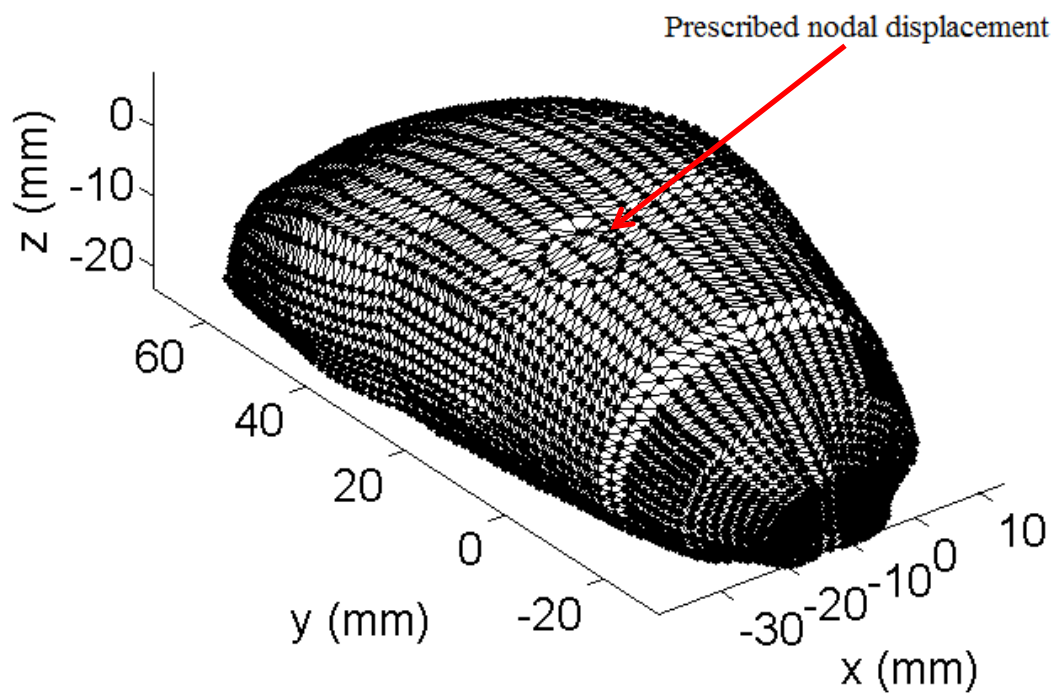

**Figure 8**: Modeling of swine brain indentation. Finite Element discretization (with 10-noded quadratic tetrahedral elements); the vertices of tetrahedras coincide with nodes of meshless discretization.

For the vast majority of the nodes, the differences in the computed deformations were under 0.1 mm and the maximum differences (occurring close to indenter edges, Figure 9) did not exceed 0.8 mm. This demonstrates that MTLED is able to replicate FEM results to good approximation despite using approximately eight times fewer nodes. As the resolution of clinical imaging systems (such as MRI) is typically not better 1 mm (often over 2 mm), these differences can be considered as negligibly small for practical purposes.

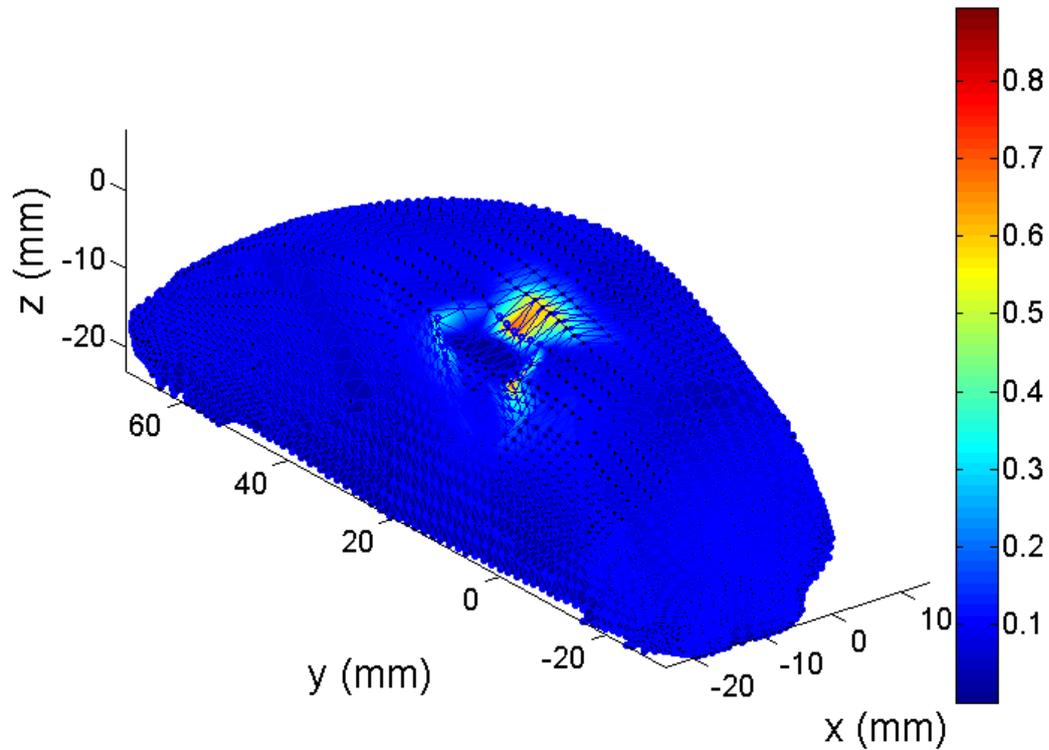

a)

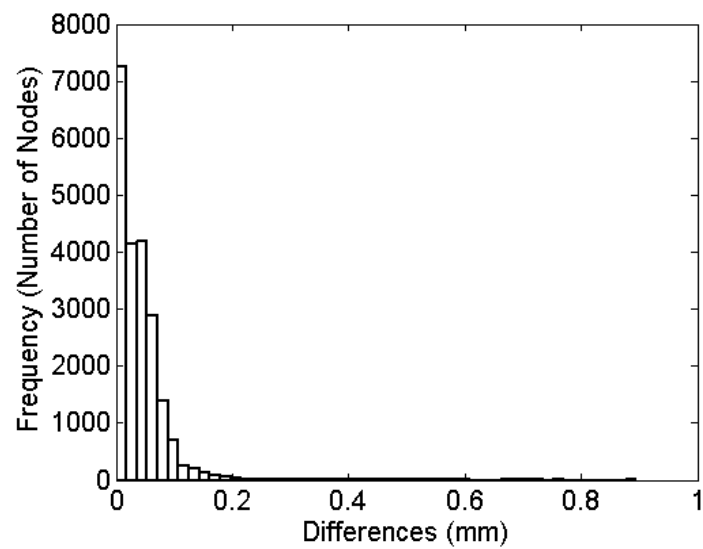

b)

**Figure 9**: Modeling of the swine brain indentation **a)** Differences in the computed deformation field between meshless (MTLED) and finite element (ABAQUS software) results (all dimensions and displacements are in mm) for the indentation depth of 5 mm, **b)** Histogram of the differences.

## 4. Experimental validation of MTLED suite of algorithms in very large deformation

In this Section we validate MTLED against extreme indentation experiments and demonstrate its applicability beyond what is possible with FEM. We consider indentation of cylindrical samples made from Sylgard 527 (Dow Corning, Midland, MI) silicone gel, down to 30% of original height. We compare the indentation force predicted with MTLED to the results computed using established non-linear algorithms available in ABAQUS (version 16.14-1) finite element code and the experimentally measured indentation force.

Figure 10 shows the experimental apparatus used to apply displacements and measure the force applied to a cylindrical gel sample (height of 17 mm and diameter of 30 mm) during indentation. The bottom surface of the sample was placed on sandpaper glued to the experimental rig base, resulting in no-slip boundary condition. The apparatus was developed and built at the Intelligent Systems for Medicine Laboratory at The University of Western Australia (Agrawal et al. 2015) (see also http://isml.ecm.uwa.edu.au/ISML/).

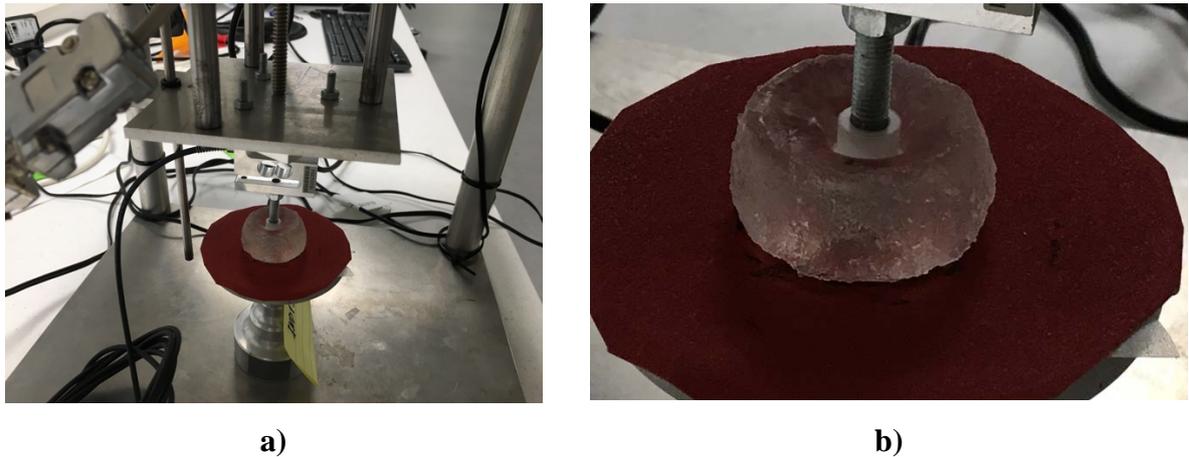

**a)**        **b)**

**Figure 10**: Indentation of cylindrical sample (diameter of 30 mm, height of 17 mm) of Sylgard 527 gel **a)** Gel sample and experimental apparatus. The indenter (hard plastic, assumed rigid) diameter is 10 mm) and **b)** close-up view of the deformed gel sample.

To appropriately model the experiment we identified the material response of Sylgard 527 gel through simple uniaxial compression. We found it to behave as Ogden-type material (Miller and Chinzei, 2002; Ogden, 1997):

$$W = \frac{2\mu_1}{a_1^2}\left(J^{-\frac{a_1}{3}}\lambda_1^{a_1} + J^{-\frac{a_1}{3}}\lambda_2^{a_1} + J^{-\frac{a_1}{3}}\lambda_3^{a_1} - 3\right) + \frac{1}{D_1}(J-1)^2 \qquad (32)$$

where $\lambda_1, \lambda_2, \lambda_3$ are the principal stretches; $a_1 = -1.1$, $\mu_1 = 643.6\ Pa$ and $D_1 = 1.2598 \times 10^{-4}\ Pa^{-1}$ are material constants.

Our MTLED model of the experiment consisted of 5,037 nodes and 25,050 integration points, Figure 11.

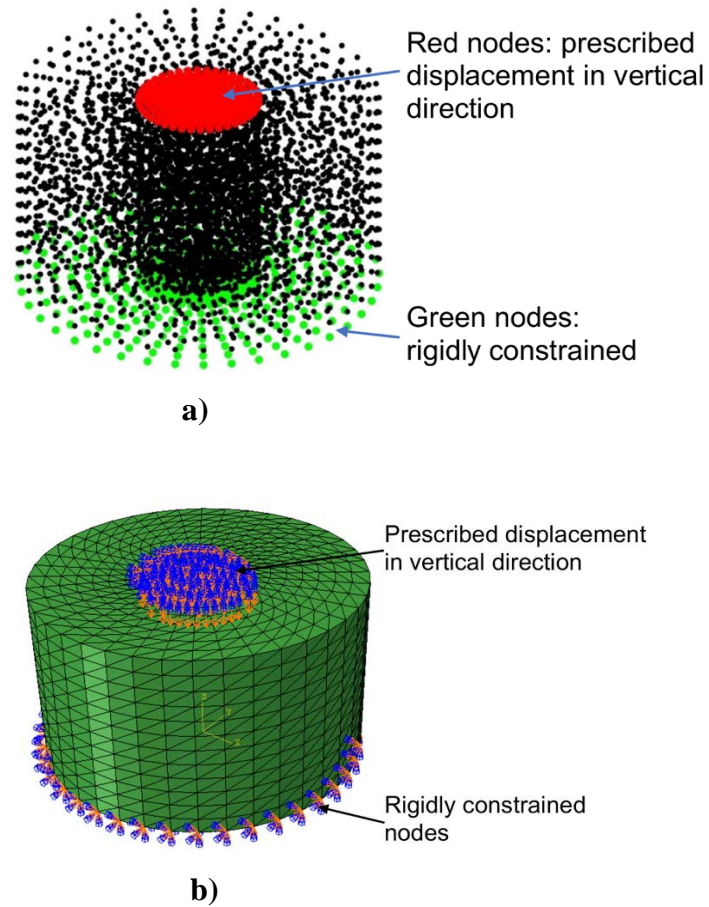

**Figure 11**: Simulation of indentation of cylindrical sample made from Sylgard 527 gel shown in Figure 10. **a)** MTLED model consisting of 5,037 nodes and 25,050 integration points in undeformed configuration. The prescribed displacement was applied at the nodes shown in red colour. The nodes shown in green were rigidly constrained. **b)** Finite element (we used a non-linear static procedure with direct solver from ABAQUS finite element code) consisting of 5037 nodes and 25050 four-noded hybrid tetrahedral elements (elements C3D4H in ABAQUS) in undeformed configuration. The boundary conditions are the same as for the MTLED model.

For comparison, we considered a non-linear (i.e. using non-linear material model eq. 32 and geometrically non-linear solution procedure) FEM model discretized with 25,050 hybrid tetrahedral elements (four-noded: element type C3D4H in ABAQUS, and ten-noded: element type C3D10H in ABAQUS) and 6,000 hybrid hexahedral elements (both eight-noded: element type C3D8H in ABAQUS, and twenty-noded: element type C3D20H in ABAQUS). The number of hexahedral elements was selected so that the number of nodes (and degrees of freedom) is close to that used in the tetrahedral meshes. The results are presented in Figures 12 and 13. MTLED and FEM give equivalent results for moderate deformations although the elements with second order (quadratic) shape functions (C3D10H and C3D20H) predict slightly lower forces than then the MTLED discretization and linear finite elements (C3D4H and C3D8H). FEM solution fails (stops to converge — we selected $10^{-9}$ iteration step as the limit) due to large element distortion at significantly lower deformations than MTLED solution.

Using the MTLED framework, we were able to obtain the stable result for the indentation depth of 12.5 mm, which is equivalent to ~75% of the height of the sample. Slight difference between the measured and computed reaction force for very large indentation depths is attributable to inadequacy of Ogden material model, eq. (32), at such extreme compressive strains.

Strains of such magnitude are common in the areas close to the contact between soft tissue and a surgical tool. For example 80% strains were seen close to the tip of a needle inserted into swine's brain (Wittek et al., 2008b). Figure 12a displays the deformed configuration of the gel sample for indentation of 12.5 mm (75% indentation) as computed by MTLED. To the best of our knowledge, it would be difficult to replicate results given in Figure 12 using any other numerical method without costly remeshing.

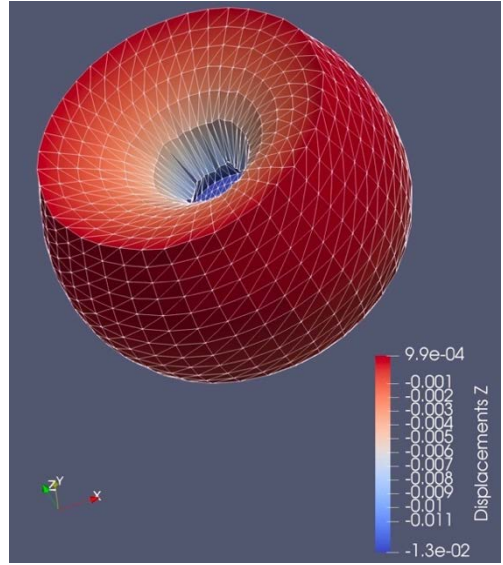

a)

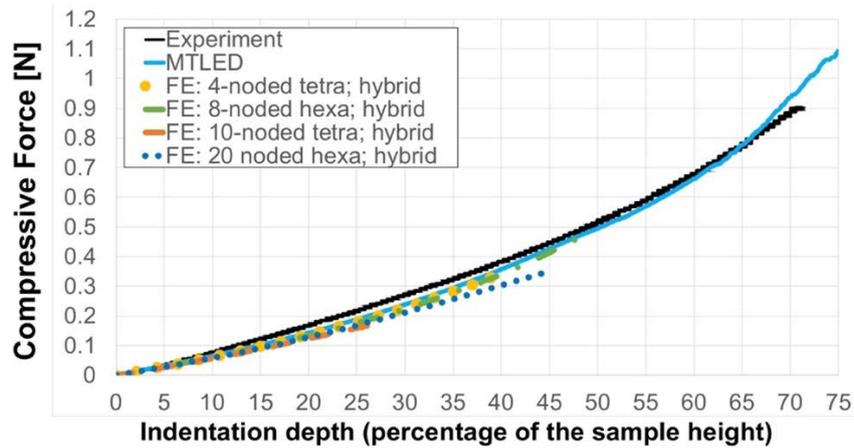

b)

**Figure 12**: Simulation of the indentation of a cylindrical sample (see Figure 10) made from Sylgard 527 silicone gel. **a)** Deformed configuration for 12.5 mm (75% of initial sample height) indentation of the gel sample predicted using MTLED. The field nodes were connected to form the triangles to visualize the deformed model surface (we used Paraview by Kitware). To the best of our knowledge result in Figure 12a would be difficult to replicate with any other numerical method without costly remeshing. The displacement scale on right-hand-side of the figure is in meters. **b)** Comparison of the force-indentation depth relationship obtained using the MTLED (blue line) and non-linear static finite element procedures (with direct solver) available in ABAQUS finite element (FE) code (yellow dotted line: 4-noded hybrid tetrahedral element, green dashed-dotted line: 8-noded hybrid hexahedral, orange dashed line: quadratic 10-noded hybrid tetrahedral elements, and blue dotted line: quadratic 20-noded hybrid hexahedral elements), and the experimental data (black solid line). 40478 steps were used in MTLED to compute the deformations for the indentation depth of 75% of the sample height. In the FE computations, the solution stopped to converge after 989 iterations (at the indentation depth around 40% of the sample height) for 4-noded hybrid tetrahedral elements; after 248 iterations (indentation depth close to 50% of the sample height) for 8-noded hexahedral hybrid elements; after 132 iterations (indentation depth less than 30% of the sample height) for 10-noded hybrid tetrahedral elements; and after 227 iterations (indentation depth of around 45% of the sample height) for 20-noded hybrid hexahedral elements. It is important to note that in 3-D non-linear solution, cost of an iteration can be as much as 4000 times higher than that of an explicit time step (Belytschko, 1976). No damping was used in the computations using MTLED.

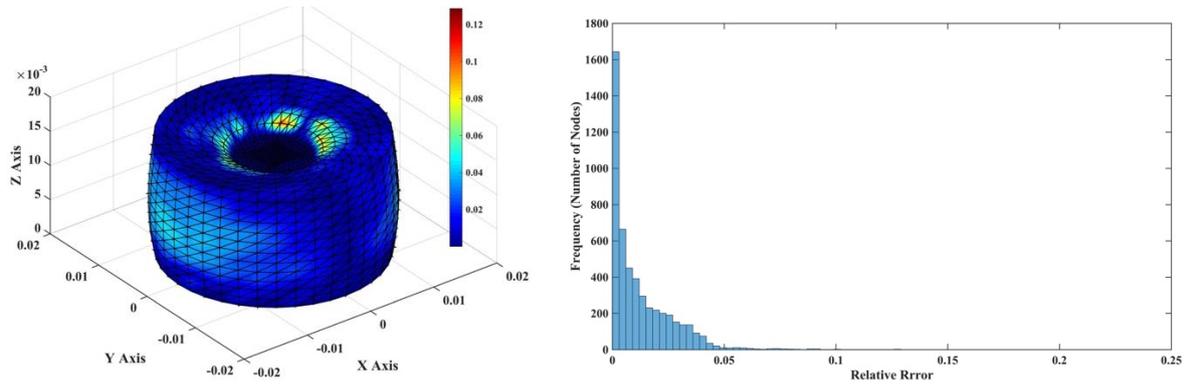

a)

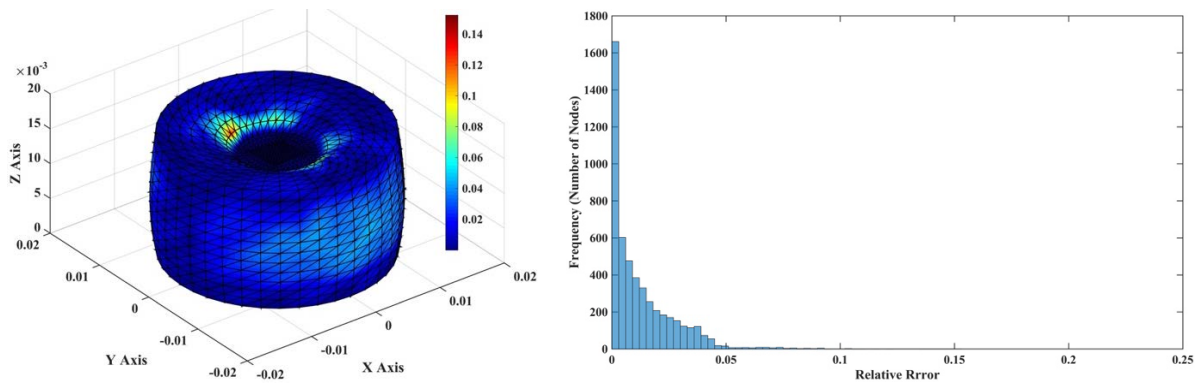

b)

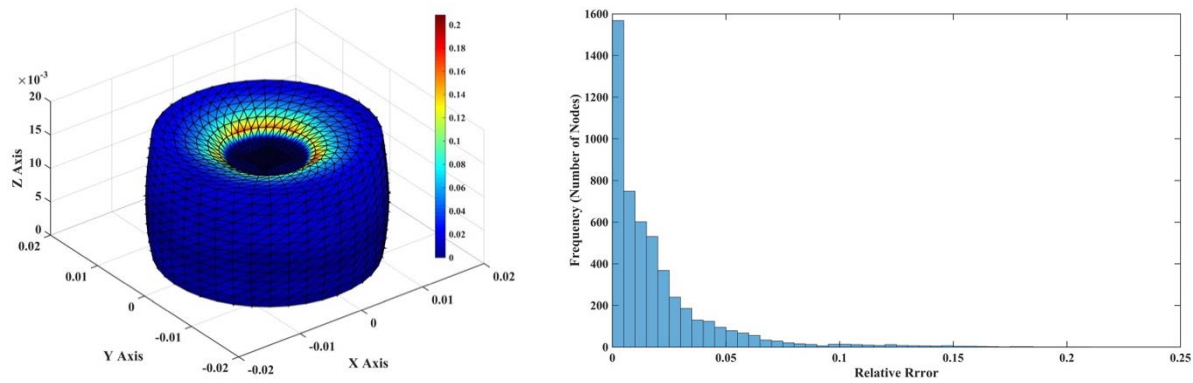

c)

**Figure 13**: Simulation of indentation of a cylinder made from Sylgard 527 gel (see Figure 11a for the experimental set-up). The figure shows differences (measured using normalized relative error (NRE)) of the computed deformation field between MTLED and ABAQUS (hybrid tetrahedral elements C3D4H) for the indentation depth of 20% of the sample height. **a)** Differences (left-hand-side column) and normalized relative error histogram (right-hand-side column) in the x-axis direction; **b)** Differences (left-hand-side column) and normalized relative error (right-hand-side column) in the y-axis direction, and **c)** Differences (left-hand-side column) and normalized relative error histogram (right-hand-side) column in the z-axis direction. The largest differences are observed in the area where the material folds into the indentation. This is the region of large strain where the differences between the finite element and meshless discretizations of the equations of solid mechanics are likely to lead to differences in predicted deformation field.

# 5. Patient-specific deformations for preoperative brain MRI to intraoperative brain CT registration.

Surgical intervention is currently the only truly curative treatment of epilepsy, and is "arguably the most underutilised of all proven effective therapeutic interventions in the field of medicine" (Engel, 2003). Functional localization is the deciding factor for many surgeries (Immonen et al., 2010). This can be dramatically improved with biomechanics-based prediction of brain tissue deformations caused by the insertion of invasive electrodes used for intracranial electro-encephalography (iEEG) (Miller et al., 2019).

These deformations can then be used to warp pre-operative MRI onto a (routinely taken) intra-operative CT (with implanted electrodes) thus providing precise geometric information about the location of the electrodes with respect to the brain anatomy as well as accurate visualisation of the deformed brain. Computational biomechanics is a powerful image registration tool (Mostayed et al., 2013a). Information about exact location of implanted electrodes and accurate visualisation of deformed brain helps localise seizure onset zones and improve surgical planning and reliability of surgery, potentially greatly increasing the number of patients undergoing this curative treatment.

Figure 14 shows a preoperative MRI and CT (with invasive electrodes implanted) of a patient undergoing epilepsy surgery planning at Boston Children's Hospital (informed consent was obtained, prior to the commencement of this study, in accordance with the BCH's Institutional Review Board).

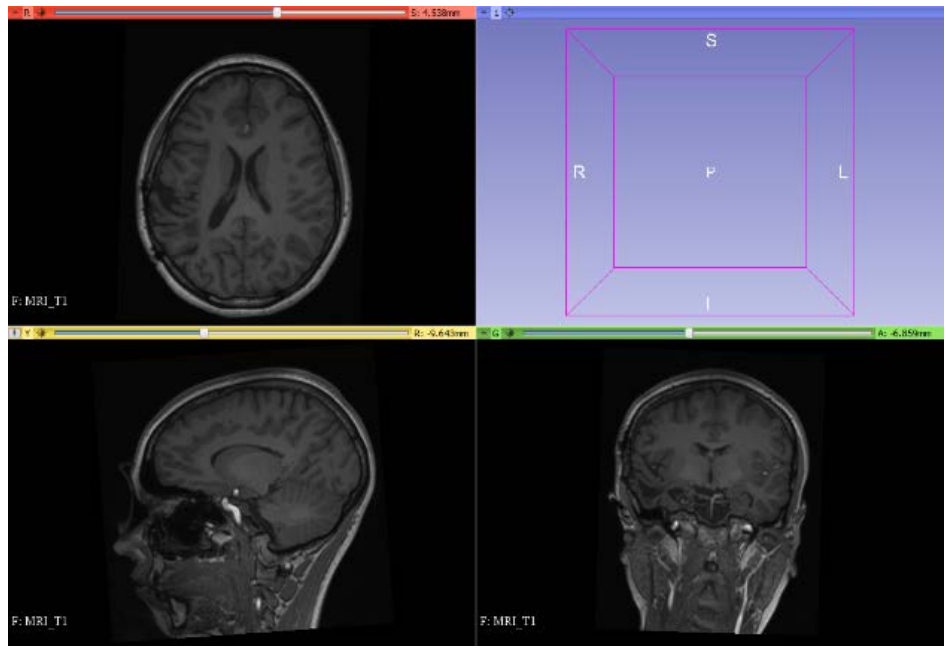

**a)**

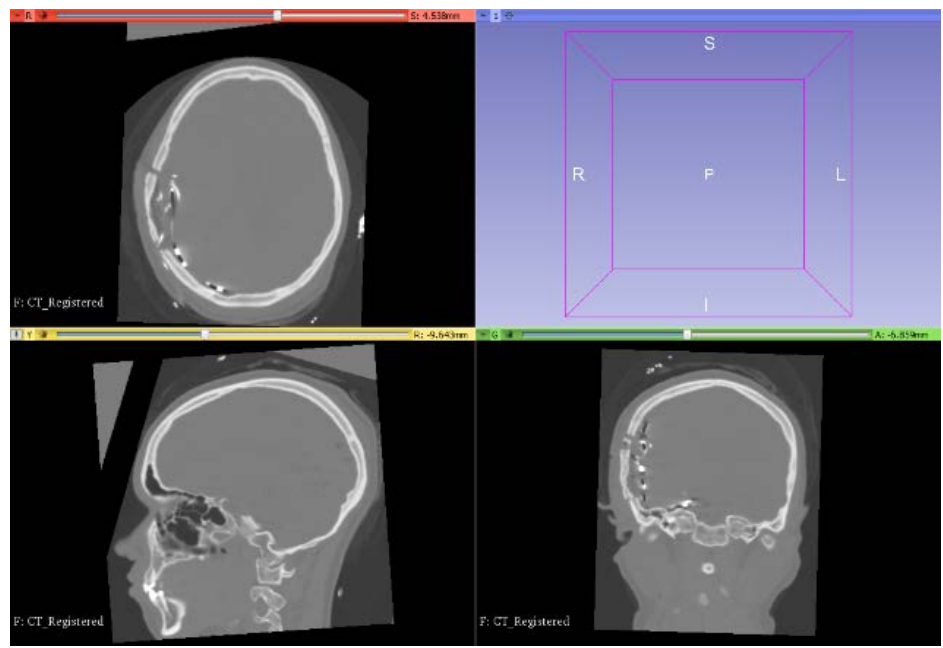

**b)**

**Figure 14**: **a)** Preoperative MRI of an epilepsy patient; **b)** CT with intracranial electrodes implanted. (Informed consent was obtained, prior to the commencement of this study, in accordance with the BCH's Institutional Review Board). Preoperative MRI to CT (with six stripes of electrodes implanted) registration is a key enabling technique in epilepsy surgery planning. Visualization performed with 3D Slicer www.slicer.org (Fedorov et al., 2012).

Our MTLED model, developed based on the preoperative MRI, consisted of 8,769 nodes and 158,678 integration points. For brain constituents we used a Neo-Hookean material model with Young's modulus allocated using Fuzzy tissue classification (3000 Pa for the brain, 100 Pa for CSF) at integration points (Li et al., 2016; Zhang et al., 2013), Figure 15. We used Poisson's ration 0.49 for all tissues. No segmentation was conducted except skull striping. The patient-specific discretization was obtained with minimal effort. This is an important advantage of MTLED over mesh-based methods.

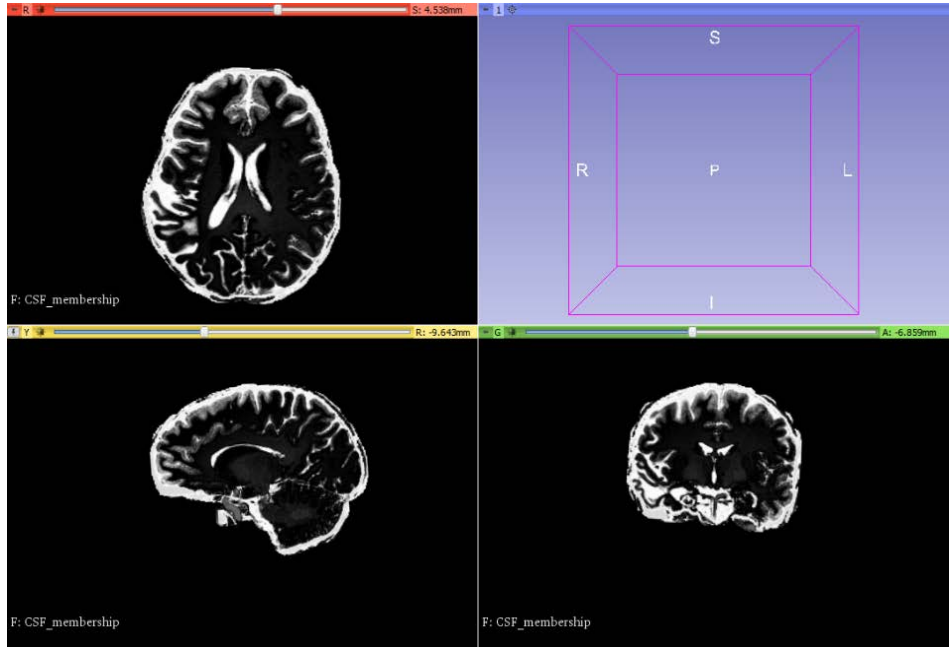

**Figure 15**: Result of automatic material property assignment (Young's modulus of 3000 Pa for the brain and 100 Pa for CSF) using fuzzy tissue classification (Li et al., 2016; Zhang et al., 2013). Brain tissue and CSF (white) were used as cluster centers. Visualization performed with 3D Slicer www.slicer.org (Fedorov et al., 2012).

Brain surface deformations due to electrode placement was obtained by projecting the positions of the electrodes as seen on CT to the brain surface segmented from preoperative MRI. These surface deformations were use as loading of the model. Maximum displacement applied was 26.7 mm.

The computed deformation field was then used to warp the preoperative MRI so that it corresponds to the brain configuration with electrodes implanted. We used our 3D Slicer module ScatteredTransform (https://www.slicer.org/wiki/Documentation/Nightly/Extensions/ScatteredTransform (Joldes, 2017)).

Figure 16 displays the computed deformation field and Figure 17 the transform used for image warping.

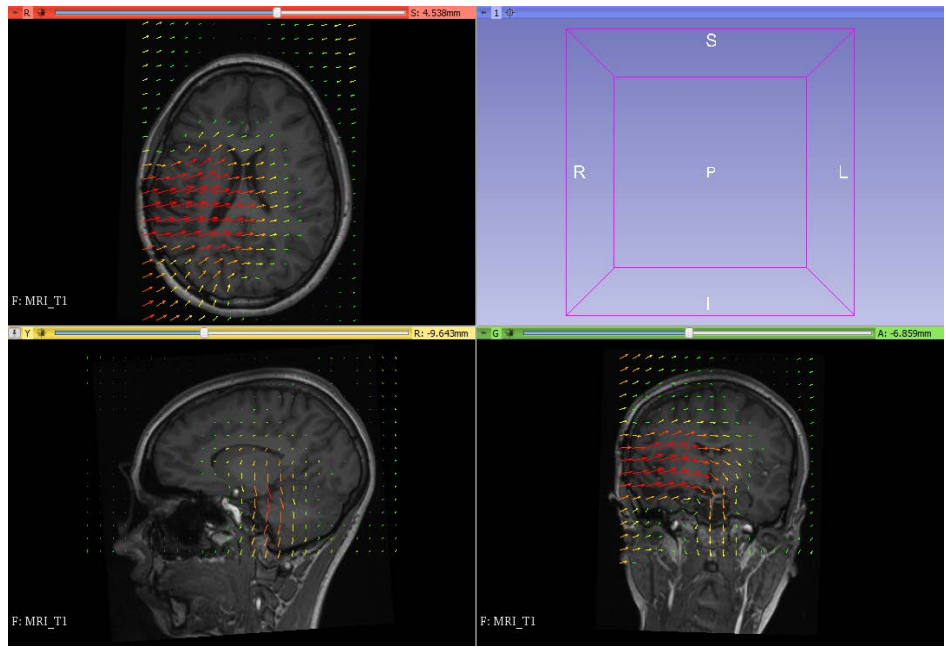

**Figure 16**: Visualization of deformation field computed by MTLED. Visualization performed with 3D Slicer www.slicer.org (Fedorov et al., 2012).

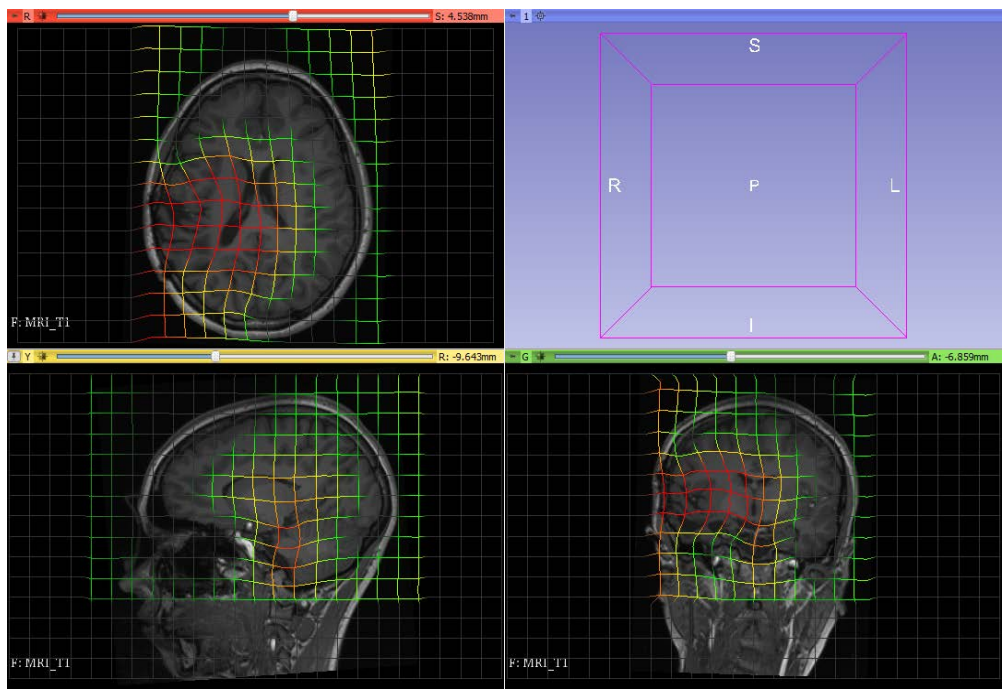

**Figure 17**: Visualization of the image transform using deformation field from Figure 16. Visualization performed with 3D Slicer www.slicer.org (Fedorov et al., 2012).

The result of registration is shown in Figure 18.

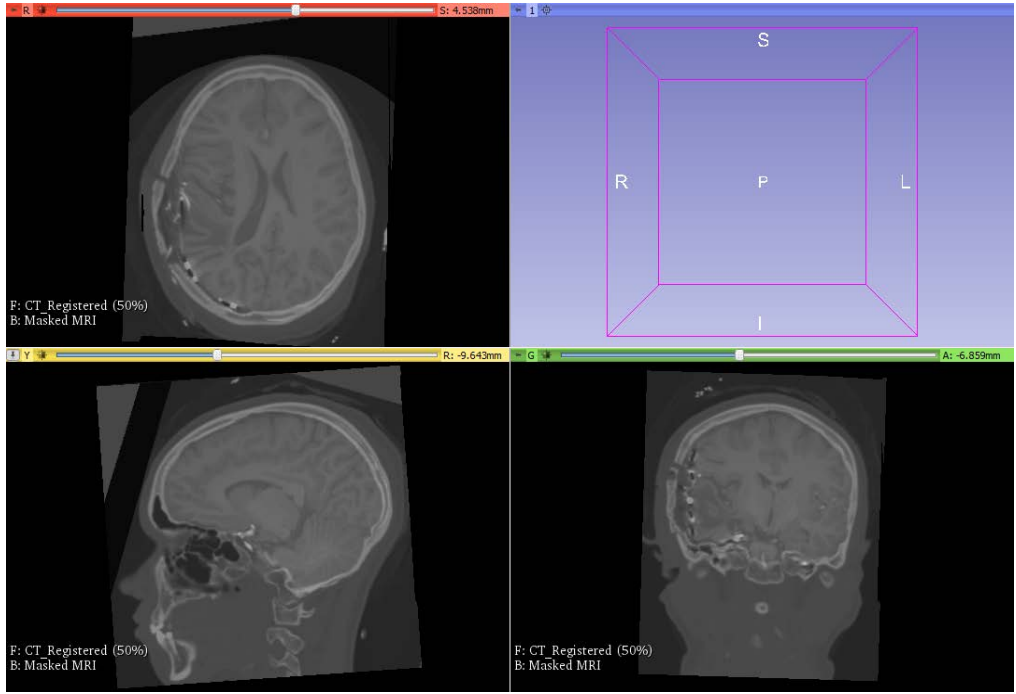

**Figure 18**: Preoperative MRI registered onto a CT with implanted intracranial electrodes (compare with the displacement field shown in Figure 16 and the transform from Figure 17). Visualization performed with 3D Slicer www.slicer.org (Fedorov et al., 2012).

Because of the simplicity of patient-specific computational grid generation the MTLED approach (and perhaps other meshless methods to be developed in the future) is better suited for integration with clinical workflows than FEM.

## 6. Discussion and Conclusions

Computational mechanics has enabled technological developments in almost every area of our lives. One of the greatest challenges for mechanists is to extend the success of computational mechanics to fields outside traditional engineering, in particular to biology, biomedical sciences, and medicine (Oden et al., 2003). By extending the surgeon's ability to plan and carry out surgical interventions more accurately and with less trauma, Computer-Integrated Surgery (CIS) systems will improve clinical outcomes and the efficiency of health care delivery. CIS systems will have a similar impact on surgery to that long since realized in Computer-Aided Design (CAD) and Computer-Integrated Manufacturing (CIM).

Robust CIS systems will contribute to the creation of a new era of personalized medicine based on patient-specific scientific computations. Sophisticated patient-specific computational models will allow optimal treatments to be tailored specifically for you. You will no longer be treated as a mean on a Gaussian distribution, or a random realization drawn from it (perhaps scaled by your body weight). The ability to rapidly build and solve such models is of paramount importance for the reliability, safe operation, and ultimately the acceptance of computational biomechanics as an integral part of Computer-Integrated Surgery (CIS) systems.

However, before this grand vision can be realized, gargantuan theoretical and technological difficulties associated with accurate and clinically practical simulation of the mechanical responses of human organs must be addressed.

First, for biomechanical computations to be practical in a clinical environment, computational grids must be obtained from standard diagnostic medical images automatically and rapidly. The current practice of patient-specific model generation involves image segmentation and finite element meshing. Both present themselves as formidable problems that are very difficult to automate. MTLED suite of algorithms presented in this paper circumvents this difficulty. Incorporation of Modified Moving Least Squares (MMLS) shape functions in MTLED increases the set of admissible nodal distributions and, as demonstrated in Section 4, allows very rapid generation of patient-specific discretization of acceptable quality.

Second, in surgical simulation interactive (haptic) rates (i.e. at least 500 Hz) are necessary for force and tactile feedback delivery (DiMaio and Salcudean, 2003). In intra-operative image registration one needs to provide a surgeon with updated images in less than 40 seconds (Warfield, 2005). To achieve these, real-time computational speeds for highly non-linear models with at least 100,000 degrees of freedom must be achieved on commodity computing hardware. MTLED uses Total Lagrangian formulation of solid mechanics and explicit time stepping. These offer a prospect of data-parallel implementation on massively parallel hardware (such as affordable GPU's) as we previously demonstrated for TLED (Joldes et al., 2010a).

Third, human soft tissues undergo very large strains in the vicinity of the contact with a surgical tool. Finite element methods are unreliable for such scenarios due to excessive element distortion, while MTLED gives reliable results for compressive strains exceeding 70%.

Fourth, surgical manipulation involves not only large deformations of soft tissues but also cutting and (often unintentional) damage. Modelling and real-time simulation of cutting, damage and propagation of discontinuities remains an unsolved and very challenging problem of computational biomechanics, however meshless methods offer advantage over finite element methods as introduction of a discontinuity in the meshless model, unlike in FE model, does not require modification of the predefined mesh but only reassignment of neighborhoods (Jin et al., 2014).

Long standing difficulties faced by many meshless methods: essential boundary condition imposition and volumetric integration of weak forms, are resolved within MTLED by incorporating the Essential Boundary Conditions Imposition for Explicit Meshless (EBCIEM) method and an adaptive numerical integration procedure that guarantees pre-specified accuracy.

We presented three numerical examples that verify that MTLED generates accurate solutions to nonlinear equations of solid mechanics governing the behavior of soft, deformable tissues. We also validate our methods against extreme indentation experiment, with the indentation depth reaching over 70% of the initial height of the sample. This result would be difficult to replicate with any other numerical method.

We also demonstrated translational benefits of MTLED using a challenging case study of predicting brain deformations due to insertion of intracranial electrodes for seizure onset zone identification for epilepsy surgery planning. Because of the ease with which an admissible discretization is generated, MTLED can be used in the clinical environment without much difficulty. The finite element method, on the other hand, is incompatible with clinical worksflows due to the requirement of high quality mesh whose generation requires formidable and labor-intensive pre-processing conducted by a specialist in image segmentation and finite element meshing.

**Acknowledgement**: The funding from the Australian Government through the Australian Research Council ARC (Discovery Project Grants DP160100714, DP1092893, and DP120100402) and National Health and Medical Research Council NHMRC (Project Grants APP1006031, APP1144519 and APP1162030) is greatly acknowledged. We thank the Raine Medical Research Foundation for supporting G. R. Joldes through a Raine Priming Grant, and the Department of Health, Western Australia, for funding G. R. Joldes through a Merit Award. H. Chowdhury and S. Agrawal are recipient of the SIRF scholarships and acknowledge the financial support of the University of Western Australia. This investigation was also supported in part by NIH grants R01 NS079788, R01 EB019483, R42 MH086984 and by a research grant from the Boston Children's Hospital Translational Research Program. We would like to thank The University of Western Australia for supporting the collaboration with Harvard Medical School through the Research Collaboration Award.